\documentclass[epj]{svjour}

\usepackage{isolatin1,graphicx,psfrag}
\usepackage[hypertex]{hyperref}

\newcommand{\pa}{p_\mathrm{A}}
\newcommand{\pb}{p_\mathrm{B}}

\begin{document}

\title{Opinion dynamics in a three-choice system}
\author{Stephan Gekle\inst{1}\thanks{E-mail: \texttt{sgekle@gmx.net}}
\and Luca Peliti\inst{2}\thanks{Associato INFN, Sezione di Napoli.
E-mail: \texttt{peliti@na.infn.it}} \and
Serge Galam\inst{1}\thanks{E-mail: \texttt{galam@ccr.jussieu.fr}}
}                     
\institute{Centre de Recherche en Épistémologie Appliquée,
CREA-École Polytechnique, CNRS UMR 7656,
1, rue Descartes, F-75005 Paris, France \and
Dipartimento di Scienze Fisiche and Unità INFM, Università
``Federico II'', Complesso Monte S. Angelo, I--80126 Napoli, Italy}
\date{Received: date / Revised version: date}
%
\abstract{
We generalize Galam's model of opinion spreading
by introducing three competing choices. At each
update, the population is randomly divided in
groups of three agents, whose members adopt
the opinion of the local majority. In the case
of a tie, the local group adopts opinion
A, B or C with probabilities $\alpha$, $\beta$
and $(1-\alpha-\beta)$ respectively. We derive
the associated phase diagrams and dynamics by
both analytical means and simulations. Polarization
is always reached within very short time scales.
We point out situations in which an initially
very small minority opinion can invade the whole system.
} 
\maketitle
\section{Introduction}
\label{intro}
In recent years, the field of Sociophysics \cite{GalShap,Gal04}
has drawn a growing interest from physicists, mostly
theoreticians, with the study of opinion dynamics as
one of the most active subjects
\cite{Gonz04,WuHub,Tessone,Slanina,Stauffer,Schweiz,Deff,Sznajd,Solomon}.
A simple model of opinion spreading with two choices was introduced
some time ago by one of the authors~\cite{Gal90}. In
this model the system will eventually reach a homogeneous state
called polarization by repeated debates. In this model,
at each time step, the population is divided at random into groups of
size $m$, and each group adopts the opinion of the local
majority. This dynamics is called \textit{randomly localized}
with a \textit{local majority} rule.
With two choices, if $m$ is odd, there are no ties and
even a slight deviation from a 50/50 initial distribution eventually
leads to the polarization of the majority opinion~\cite{Gal90,Gal02}.

Here we investigate the richer dynamics provided by an $m=3$
system with \textit{three} possible choices, which we shall label
by A, B, and C. This leads to the possibility of ties, when
each member of the group has a different opinion~\cite{Gal91,Gal00}.
We introduce the probabilities $\alpha$, $\beta$ and $(1-\alpha-\beta)$
of resolving the tie in favor of A, B and C respectively.
An alternative proposal for the discussion dynamics has been
considered in ref.~\cite{Aerts}, where quantities analogous
to $\alpha$ and $\beta$ allow for a contextual interpretation. The
quantum nature of these models could reflect
more precisely the ``reduction process''
by which a discussant takes up an opinion.
However, in this work the focus lies in the reciprocal interference
of opinions on three different questions, while here we are interested in
a situation in which opinions concerning a single question can take up
three values.

In our model we find that
polarization is always reached in a short time, and that it is
possible to summarize the behavior of the system (for given
values of $\alpha$ and $\beta$) by associating to each initial
opinion distribution the value of the opinion which eventually
prevails. We can thus draw phase diagrams, exhibit phase separation
lines and fixed points. In particular it is found that in some circumstances
an initially very small minority opinion can invade the whole system.
While the phase boundaries are sharp in the limit of
large populations, simulations show that they become blurred
when the population is small,
since starting from the same initial condition may lead to the
polarization of different opinions. The polarization time also
exhibits larger fluctuations, always remaining very small.

In Section~\ref{equations:sec}, after a brief review of
the standard Galam model
with two choices~\cite{Gal90}, we define the model with three choices
and we derive the basic evolution equations. The fixed points and flows
of these equations are discussed in Section~\ref{flow:sec}. These analytical
predictions are compared against numerical simulations in
Section~\ref{simul:sec}. Section~\ref{concl:sec} contains some
concluding remarks and proposes some research possibilities.

\section{The model}
\label{equations:sec}
We first describe the randomly localized dynamics with local majority
rule for the standard Galam model~\cite{Gal90,Gal02} with two
choices, and we then introduce our present model with three
choices.

We consider a population of $N$ agents, each of which can have
opinion A or opinion B. At each time step, the whole population
is divided into groups of size $m$, so that each agent belongs to
one and only one group. Discussions take place, and at the end
of the discussion all the members of the group adopt the opinion
of the local majority. The process is iterated until a stable
situation is reached.

With $m=3$, ties are ruled out and we can write outright the
evolution equation for the fraction $\pa$ of agents
with opinion $A$~\cite{Gal02}:
\begin{equation}
\pa(t+1)=\pa^3(t)+3\pa^2(t)(1-\pa(t)).
\label{Gal:eq}
\end{equation}
This evolution exhibits the two stable fixed points
at $\pa=0$, $\pa=1$, and an unstable
fixed point at $\pa=\frac12$, which defines
the phase boundary. This means that an initial condition slightly
different form 50/50 will always lead to a complete polarization
of the system, since the flow is drawn towards one of the
stable fixed points.

When there are three choices, A, B and C, there is the possibility of
a tie in which each agent has a different opinion. We assume that
the tie is always resolved in favor of one of the opinions, and
we introduce the probabilities $\alpha$, $\beta$ and
$\gamma=1-\alpha-\beta$ that
the winning opinion is A, B, or C respectively.
Thus, e.g., the three members of a group will come out
with opinion A if one of these cases applies:
\begin{enumerate}
\item They all already have opinion A, which happens with probability
$\pa^3$;
\item Two of them have opinion A, which happens with
probability $3 \pa^2(1-\pa)$;
\item There is a tie resolved in favor of A, which happens
with probability
$6\alpha \pa\pb(1-\pa-\pb)$.
\end{enumerate}
A similar analysis can be made for $\pb$.
We consider the large population limit, in which sampling
fluctuations can be neglected.
We thus obtain the evolution equations for the fractions
$\pa$, $\pb$:
\begin{eqnarray}
\pa(t+1)&=&\pa^3(t)+3\pa^2(t)(1-\pa(t))
\nonumber\\
&&{}+6\alpha \pa(t)\pb(t)
(1-\pa(t)-\pb(t));
\label{eva:eq}\\
\pb(t+1)&=&\pb^3(t)+3\pb^2(t)(1-\pb(t))
\nonumber\\
&&{}+6\beta \pa(t)\pb(t)
(1-\pa(t)-\pb(t)).\label{evb:eq}
\end{eqnarray}

The state of the system can be conveniently represented
by a point in an equilateral triangle. Let $\vec A$, $\vec B$ and
$\vec C$ be the vertices of such a triangle, then the
state defined by the probabilities $\pa$,
$\pb$ and $p_\mathrm{C}=1-\pa-\pb$
is represented by the point
\begin{eqnarray}
\vec p&=&\pa\vec A+
p_\mathrm{b}\vec B+p_\mathrm{C}\vec C\nonumber\\
&=&\vec C+\pa \overline{\mathrm{CA}}+\pb
\overline{\mathrm{CB}},
\end{eqnarray}
where $\overline{\mathrm{CA}}$ and $\overline{\mathrm{CB}}$
are the vectors leading from $\vec C$ to $\vec A$ and $\vec B$
respectively. The vertices of the triangle represent the
polarized states. Inspection of eqs.~(\ref{eva:eq},\ref{evb:eq})
(and common sense) shows that these states are fixed and stable.
Each vertex thus commands a nonempty attraction basin.
Indeed, one can see that if, e.g., $\frac{1}{2}<\pa(t)<1$,
one has $\pa(t+1)>\pa(t)$. Thus the attraction basin
of $\vec A$ contains at least the region $\pa>\frac{1}{2}$.

\section{Fixed points and flow diagrams}
\label{flow:sec}
\begin{figure}[htb]
\begin{center}
\includegraphics[width=6cm]{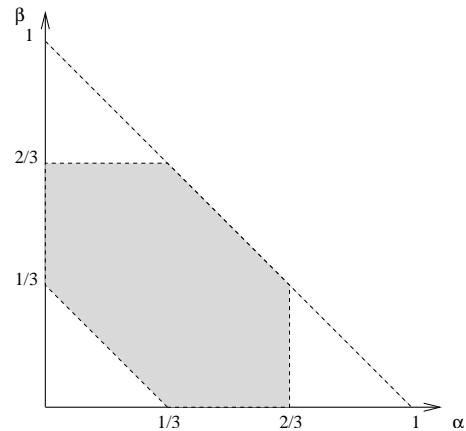}
\end{center}
\caption{The region in the $(\alpha,\beta)$ plane in which the
inner fixed point G appears is dashed.}
\label{innerpt:fig}
\end{figure}

Equations (\ref{eva:eq},\ref{evb:eq}) exhibit the following
fixed points:
\begin{enumerate}
\item The trivial fixed points A, B and C, corresponding
to the polarized states;
\item Since, when one of the three fractions $\pa$,
$\pb$ or $p_\mathrm{C}$ vanishes, the evolution
equations reduce to the one describing the two-choice Galam
model (eq.~(\ref{Gal:eq}), there are the corresponding
fixed points D ($\pa=\frac12$, $\pb=\frac12$),
E ($\pb=\frac12$, $p_\mathrm{C}=\frac12$), and
F ($p_\mathrm{C}=\frac12$, $\pa=\frac12$) which
lie on the triangle sides.
\item Moreover a seventh fixed point G, in which
all three fractions are nonzero, may appear in
the DEF triangle. For this point to
appear, it is necessary that
none of the probabilities $\alpha$, $\beta$
and $\gamma=1-\alpha-\beta$ is larger than $\frac{2}{3}$.
Thus the region in the $(\alpha,\beta)$ plane in which this
G point appears is the one dashed in fig.~\ref{innerpt:fig}.
\end{enumerate}

\begin{figure*}[htp]
\begin{center}
\includegraphics[width=6cm]{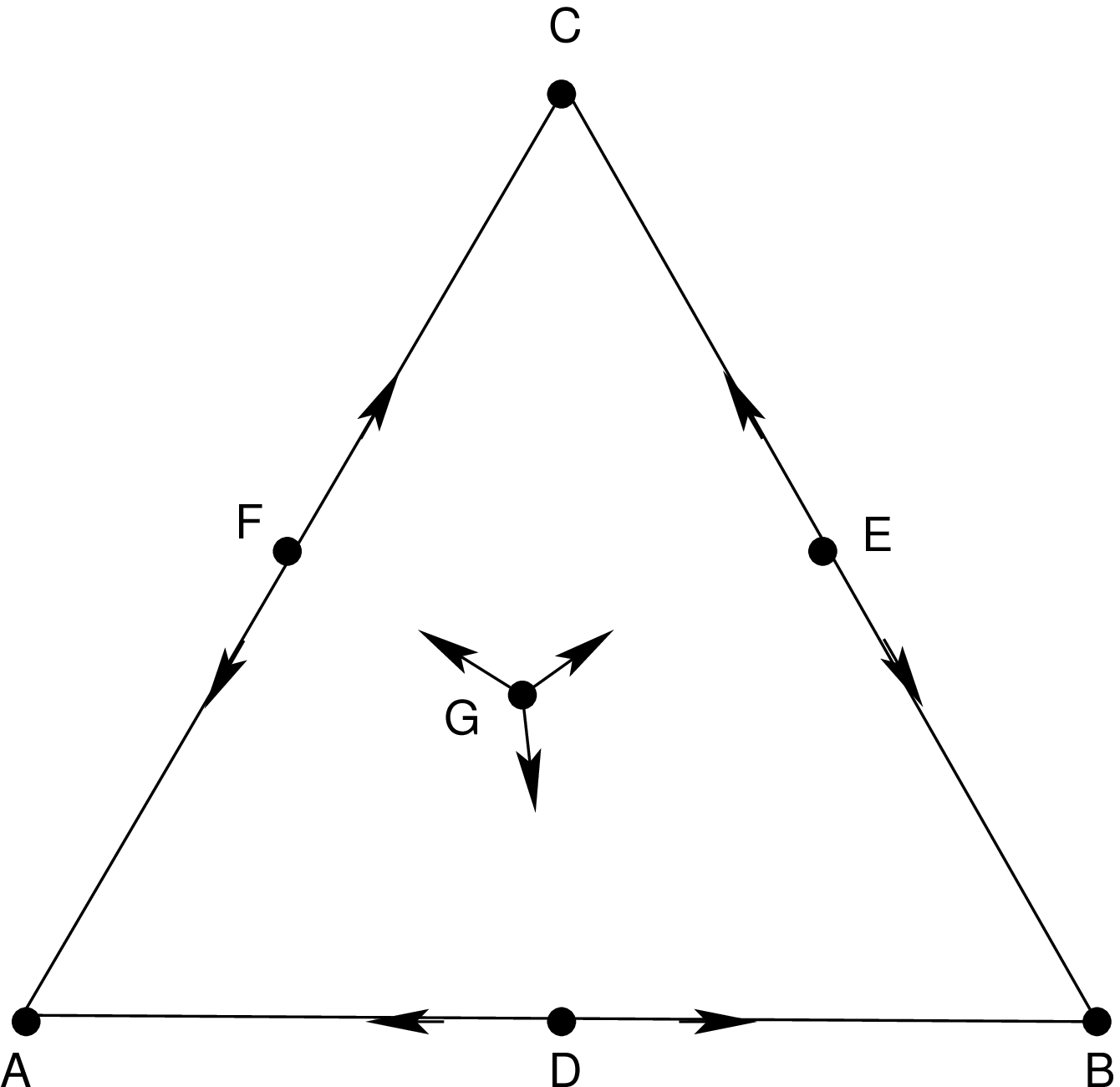}
\hspace{0.5cm}
\includegraphics[width=6cm]{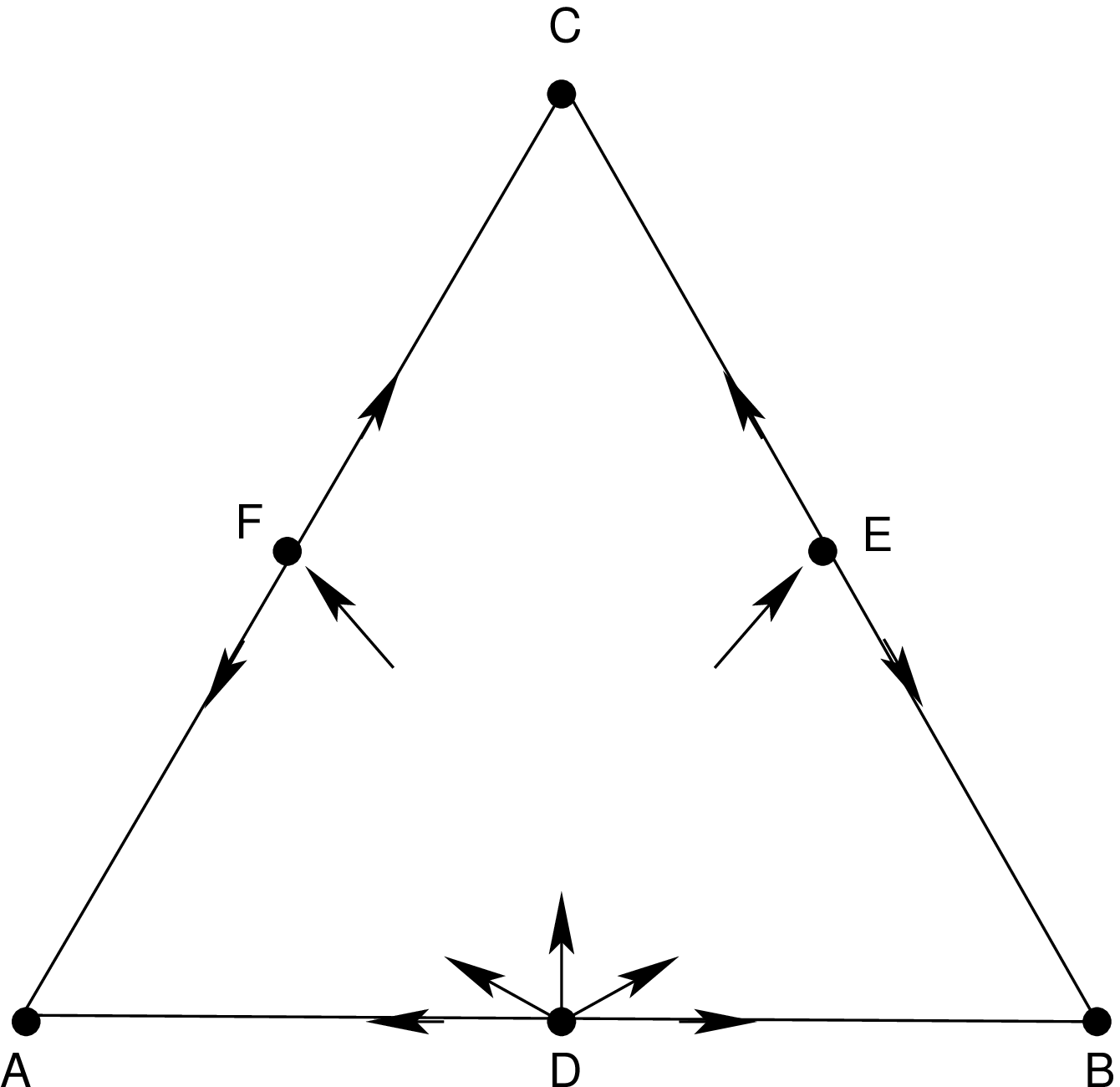}
\end{center}
\caption{Fixed points and basic structure of
the flow. Left: Case I, in which $(\alpha,\beta)$ lies inside
of the dashed region in fig.~\ref{innerpt:fig}. Right: Case II, in which
$(\alpha,\beta)$ lies outside that region. In the case shown,
$\gamma=1-\alpha-\beta>\frac{2}{3}$.}
\label{triang:fig}
\end{figure*}

The stability of the fixed points can as usual be discussed
by linearizing the evolution equations around them. If we denote
by $J$ the Jacobian matrix of the transformations
(\ref{eva:eq},\ref{evb:eq}), a fixed point is stable
if all the eigenvalues of $J$ are smaller than 1 in absolute value.
One can easily check that:
\begin{enumerate}
\item The three trivial fixed points A, B and C have all
eigenvalues equal to 0;
\item The three side points D, E and F, have one eigenvalue
equal to $\frac{3}{2}$, corresponding to the eigenvector
lying along the side of the triangle, whereas
the other one is given by
\begin{equation}
\frac{3}{2}(1-\alpha-\beta) \mbox{ for D},\quad
\frac{3}{2}\alpha \mbox{ for E},\quad \mbox{and  }
\frac{3}{2}\beta \mbox{ for F}.
\end{equation}
We see that the direction towards the interior of the triangle
becomes unstable if the probability parameter associated with
the opposite vertex is larger than $\frac{2}{3}$. In this
situation, the inner fixed point does not exist.
\item The inner fixed point, if it exists, has both
eigenvalues larger than 1 in modulus.
\end{enumerate}
The basic structure of the flow is thus represented in fig.~\ref{triang:fig}.
The first applies to the case in which there is
the inner fixed point, G, (Case~I) and the other to
the case in which it does not exist (Case~II).
It is clear from this figure that, in Case~I, the
attraction basins of the stable fixed points meet two by
two at the fixed points D, E and F, and all three in the
inner point G. In Case~II, on the other hand, the attraction
basin of the point corresponding to the largest favorable
probability ($\alpha$ for A, etc.) will separate the basins
of the other fixed points at the side fixed point situated
midway between the two.

\begin{figure*}[htb]
\begin{center}
\includegraphics[width=6cm]{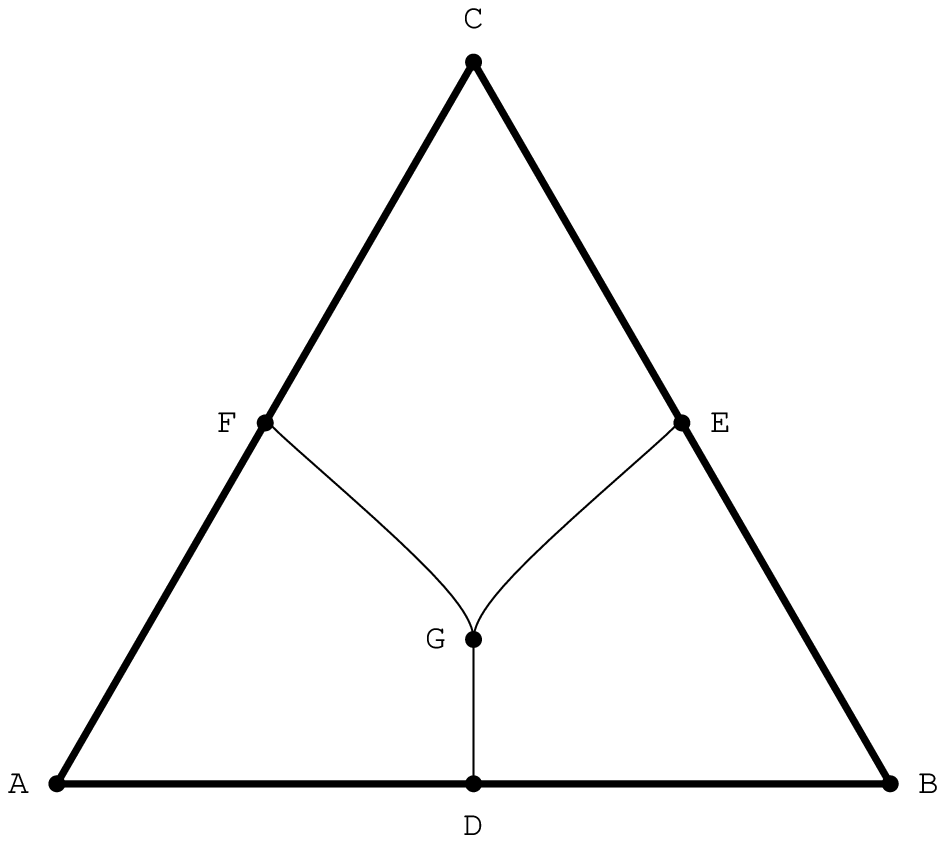}
\hspace{0.5cm}
\includegraphics[width=6cm]{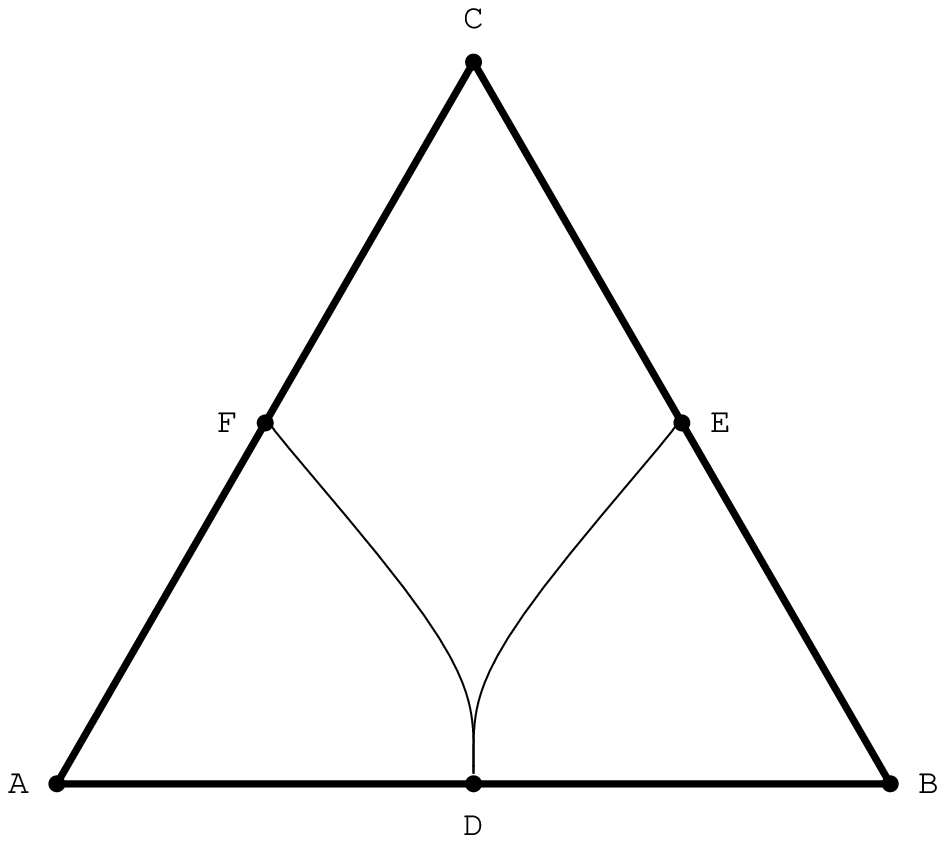}
\end{center}
\caption{Phase diagrams for the dynamics: Left:
$\alpha=\beta=\frac{1}{4}$ (Case~I). Right: $\alpha=\beta=\frac{1}{6}$
(Case~II).}
\label{phasediag:fig}
\end{figure*}

\begin{figure*}[htp]
\begin{center}
\includegraphics[width=6cm]{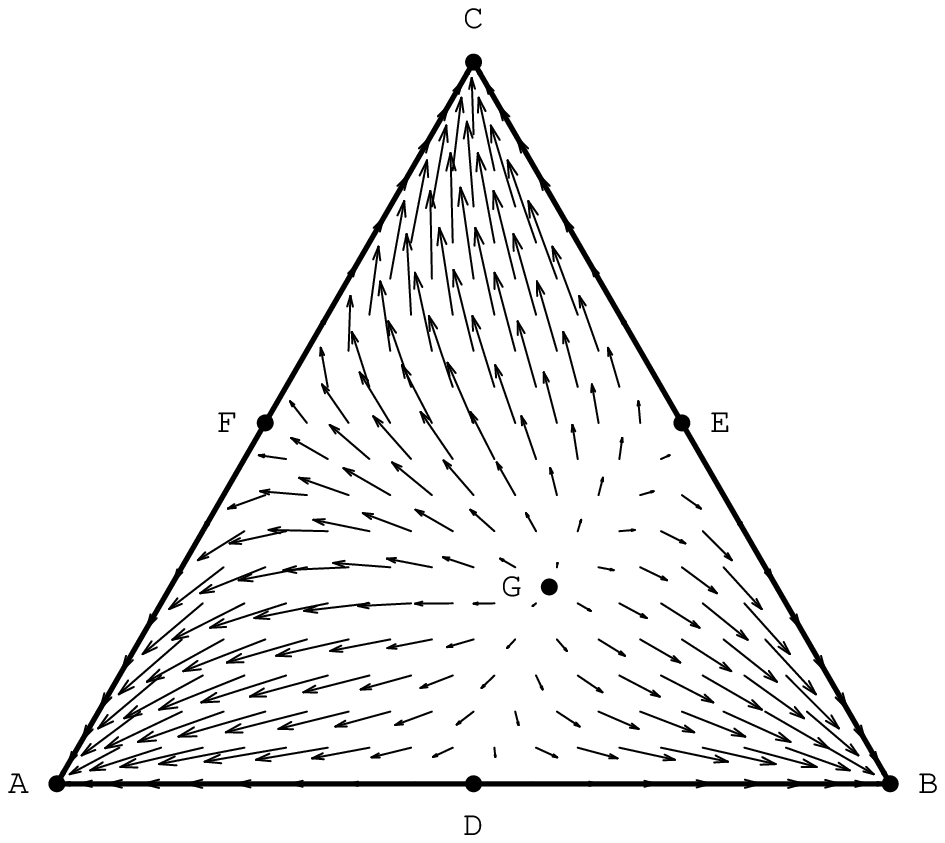}
\hspace{0.5cm}
\includegraphics[width=6cm]{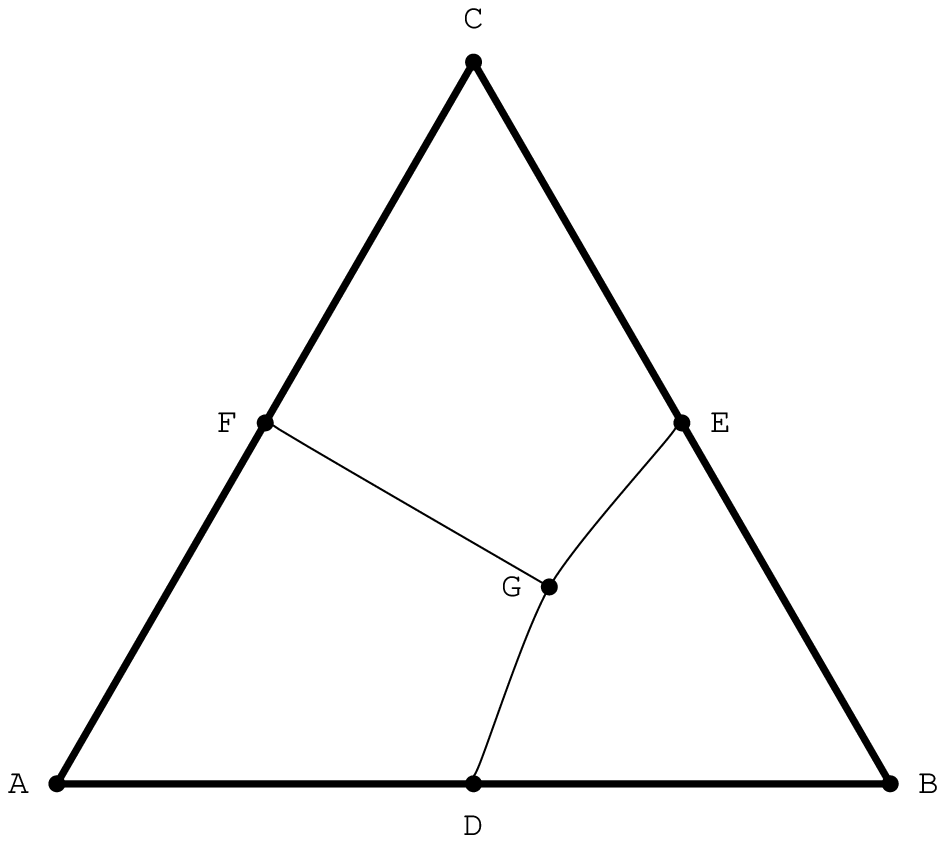}
\end{center}
\caption{Flow, fixed points and attraction
basins for $\alpha=4/9$, $\beta=1/9$ (Case~I).
Left: Flow and fixed points. Right: Boundaries
of the attraction basin.}
\label{918:fig}
\end{figure*}

\begin{figure*}[htp]
\begin{center}
\includegraphics[width=6cm]{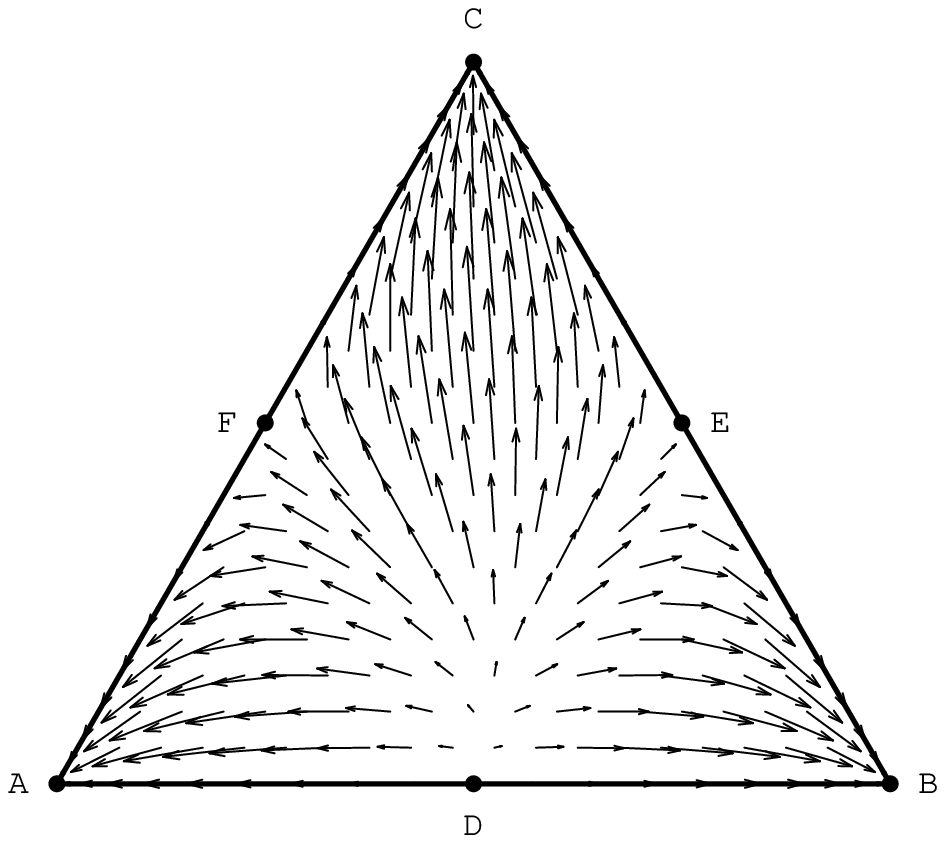}
\hspace{0.5cm}
\includegraphics[width=6cm]{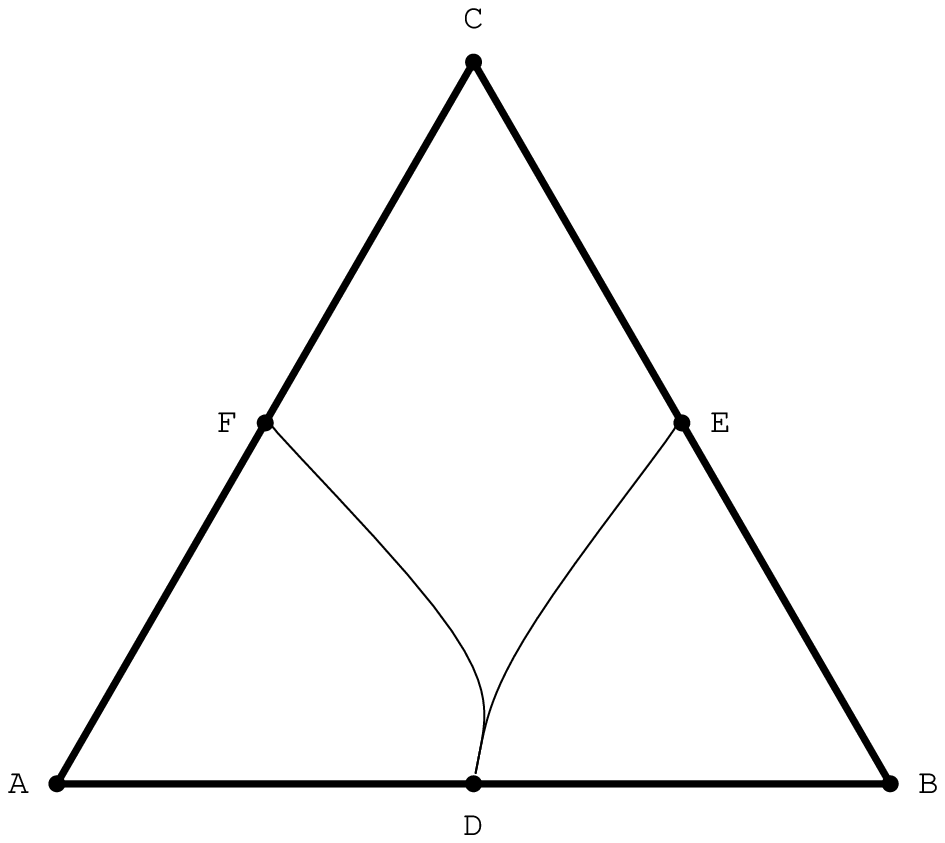}
\end{center}
\caption{Flow, fixed points and attraction
basins for $\alpha=2/9$, $\beta=1/9$ (Case~II).
Left: Flow and fixed points. Right: Boundaries
of the attraction basin.}
\label{1318:fig}
\end{figure*}

The phase boundaries can be conveniently located by
iterating the inverse of eqs.~(\ref{eva:eq},\ref{evb:eq}),
starting close to the side fixed points D, E, or F. The iterations
converge to the inner fixed point G, if it exists, or to
the other relevant side fixed point. As an example,
we show in fig.~\ref{phasediag:fig} the phase diagram obtained
for $\alpha=\beta=\frac{1}{4}$ and $\frac{1}{6}$ respectively.
It is clear that the attraction basins of the side
fixed points reduce to stretches of the phase boundaries:
the one connecting each point to G, in Case~I, or to the
unstable side fixed point (like D in fig.~\ref{phasediag:fig}, right),
in Case~II. In figures~\ref{918:fig} and \ref{1318:fig}
we show respectively the flow and the phase diagrams for
$\alpha=1/2$, $\beta=1/3$ (Case~I) and for $\alpha=2/9$, $\beta=1/9$.
We can conclude that in Case~II, close to the unstable side
fixed point, there is the possibility that the ``preferred'' opinion
(the one with the largest probability of resolving a tie) can invade
the system even if it is initially supported by a very small minority.
The width
of the sector in the $\vec p=(\pa,\pb)$ plane where
this can happen rapidly vanishes as $\vec p$ approaches the
side of the triangle, since the two boundary lines are tangent
to each other (unless $\alpha=\beta=0$).
The sector, however, opens up more and more rapidly
as the third opinion becomes more and more preferred.

\begin{figure*}
\begin{center}
\includegraphics[width=4cm]{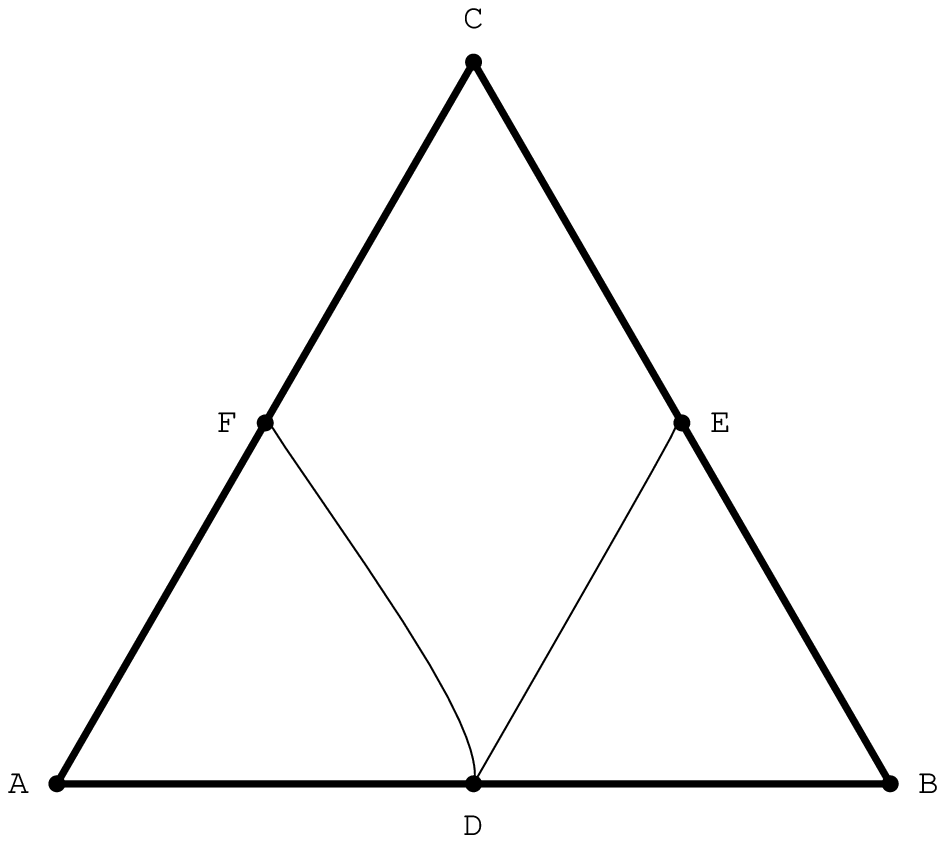}
\hspace{0.2cm}
\includegraphics[width=4cm]{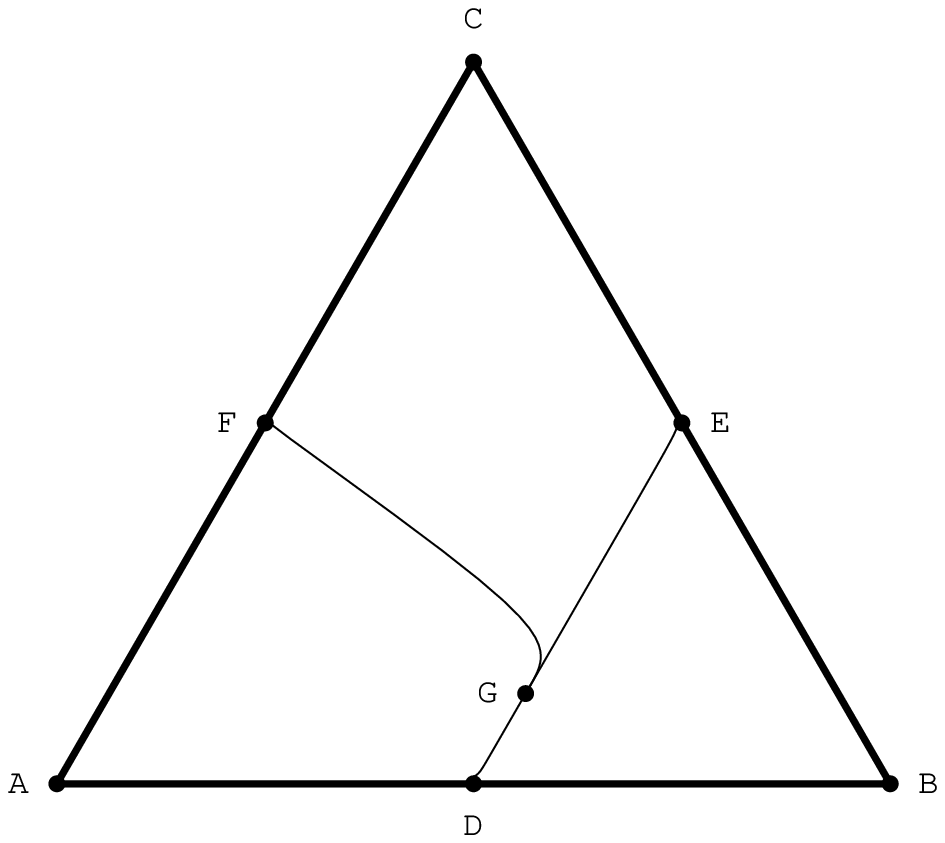}
\hspace{0.2cm}
\includegraphics[width=4cm]{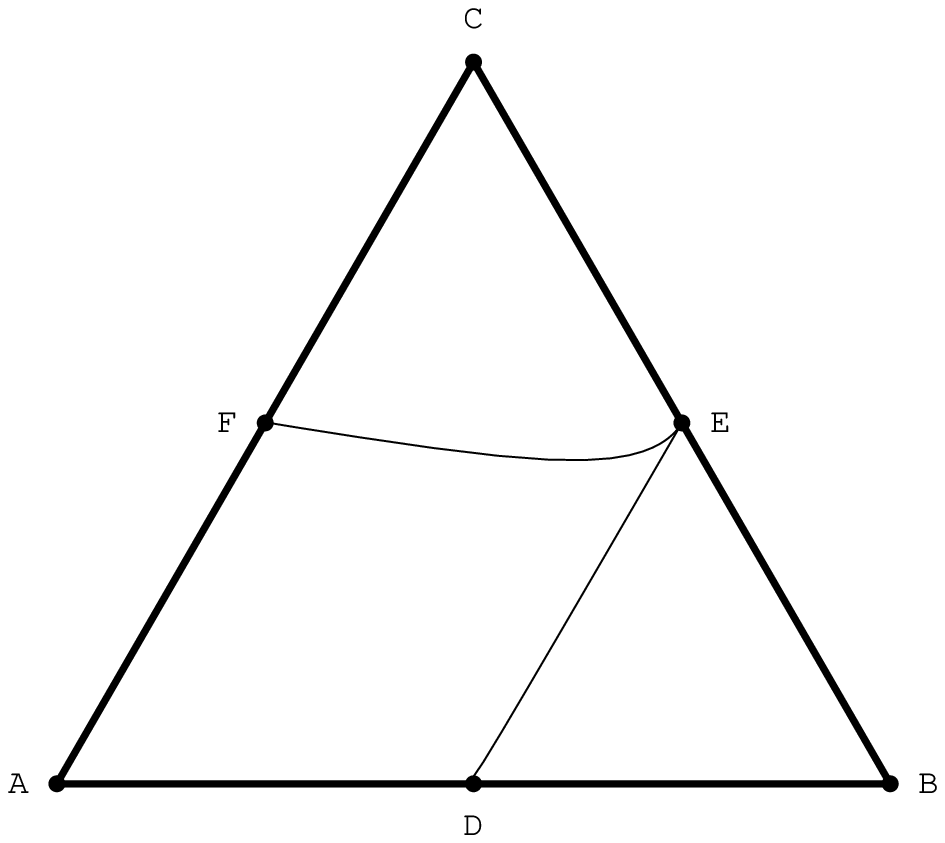}
\end{center}
\caption{Phase diagrams with $\beta=0$. The straight line
DE ($\pb=\frac{1}{2}$) is conserved by the flow.
Left: $\alpha=\frac{1}{10}$. Center: $\alpha=\frac{5}{12};$
Right: $\alpha=\frac{5}{6}$.}
\label{zerobeta:fig}
\end{figure*}

If one of the three probabilities (e.g., $\beta$) vanishes,
the straight line $\pb=\frac{1}{2}$,
connecting the side fixed points
D and E on either side of the corresponding vertex B,
is conserved by the flow and defines
the boundary of the attraction basin for it. This means that
the ``unpreferred'' opinion can invade
only if it is initially shared by a majority.
For $\alpha<\frac{1}{3}$
the fixed point E is stable along this line. For
$\frac{1}{3}<\alpha<\frac{2}{3}$ the inner fixed point G appears on
the line, and both side fixed points D and E are
partially stable. Finally, if $\alpha>\frac{2}{3}$, the
fixed point D becomes partially stable, while E becomes
totally unstable. The different cases are represented in
fig.~\ref{zerobeta:fig}.

\section{Simulations}
\label{simul:sec}

\subsection{Phase diagram}
\begin{figure*}
\begin{center}
\begin{tabular}{ccccc}
\includegraphics[width=2.6cm]{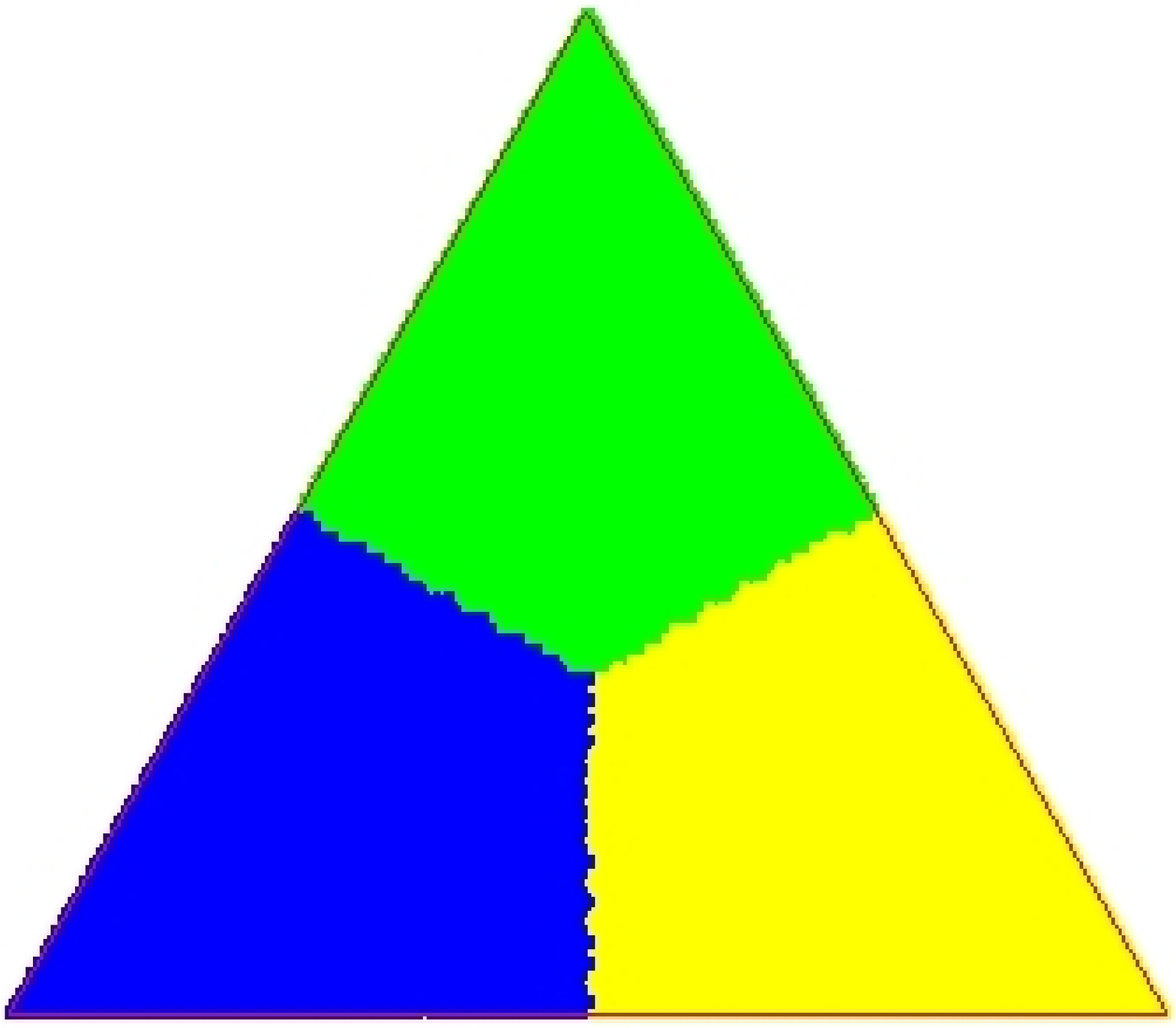} &
\includegraphics[width=2.6cm]{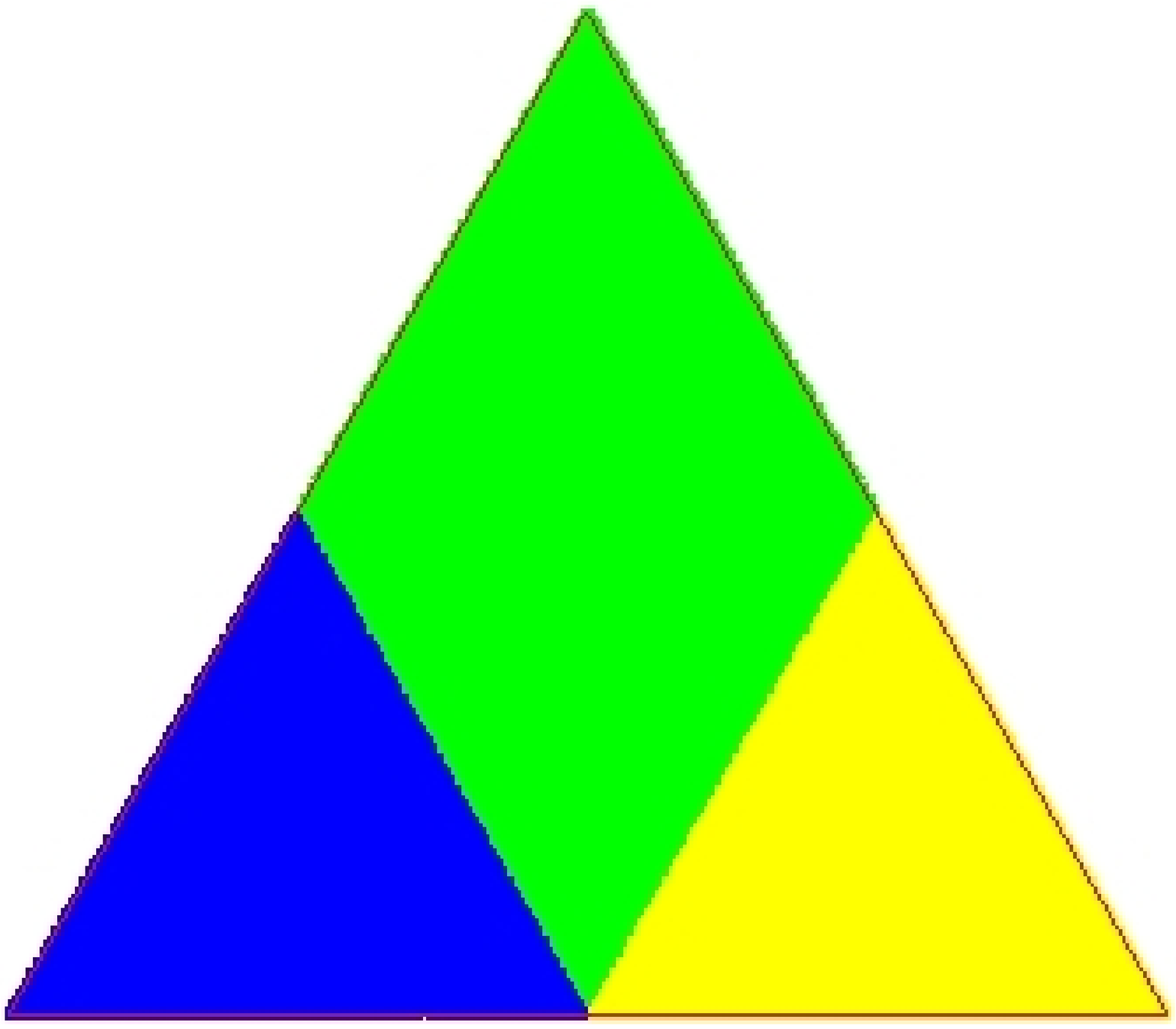}&
\includegraphics[width=2.6cm]{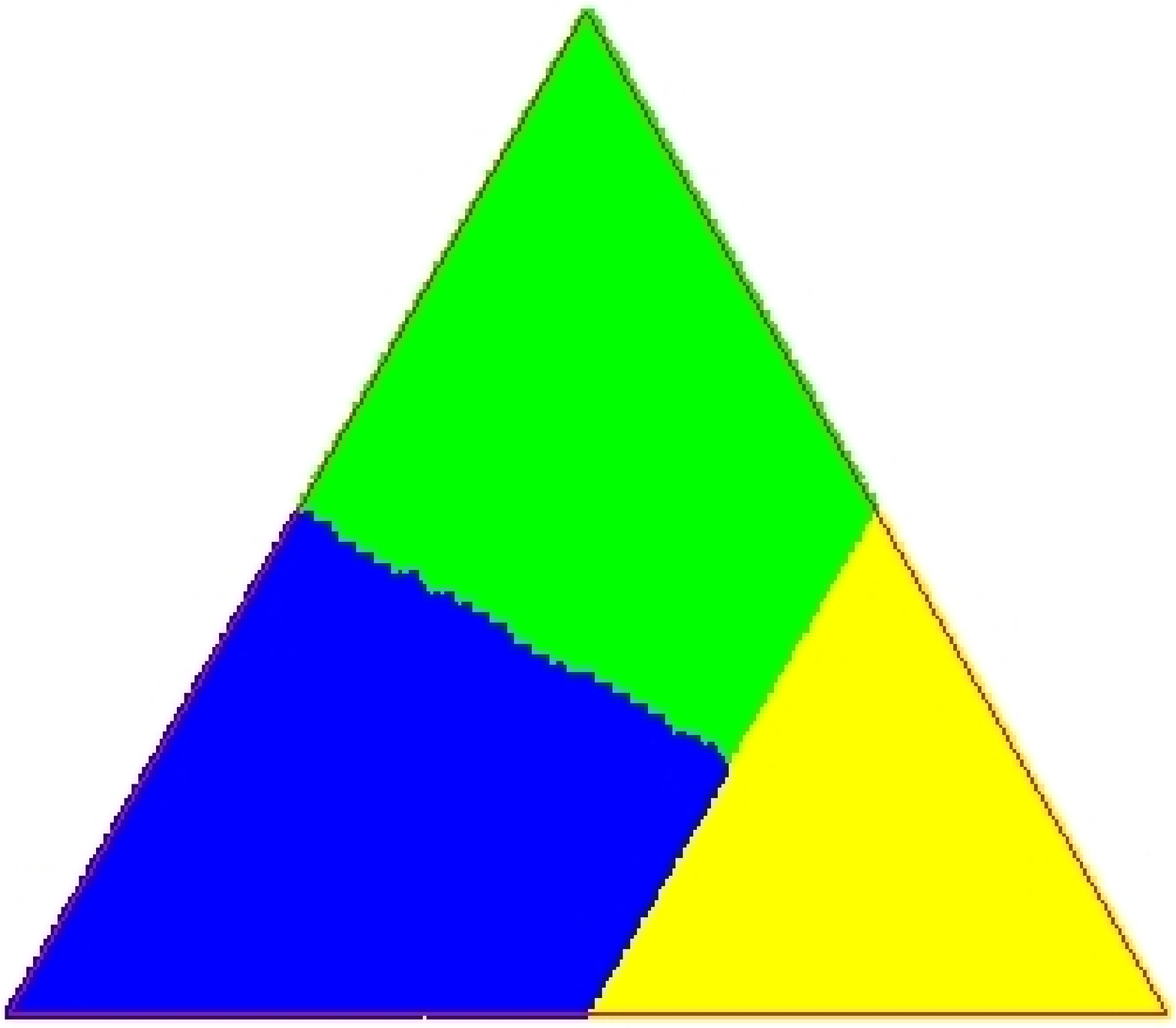}&
\includegraphics[width=2.6cm]{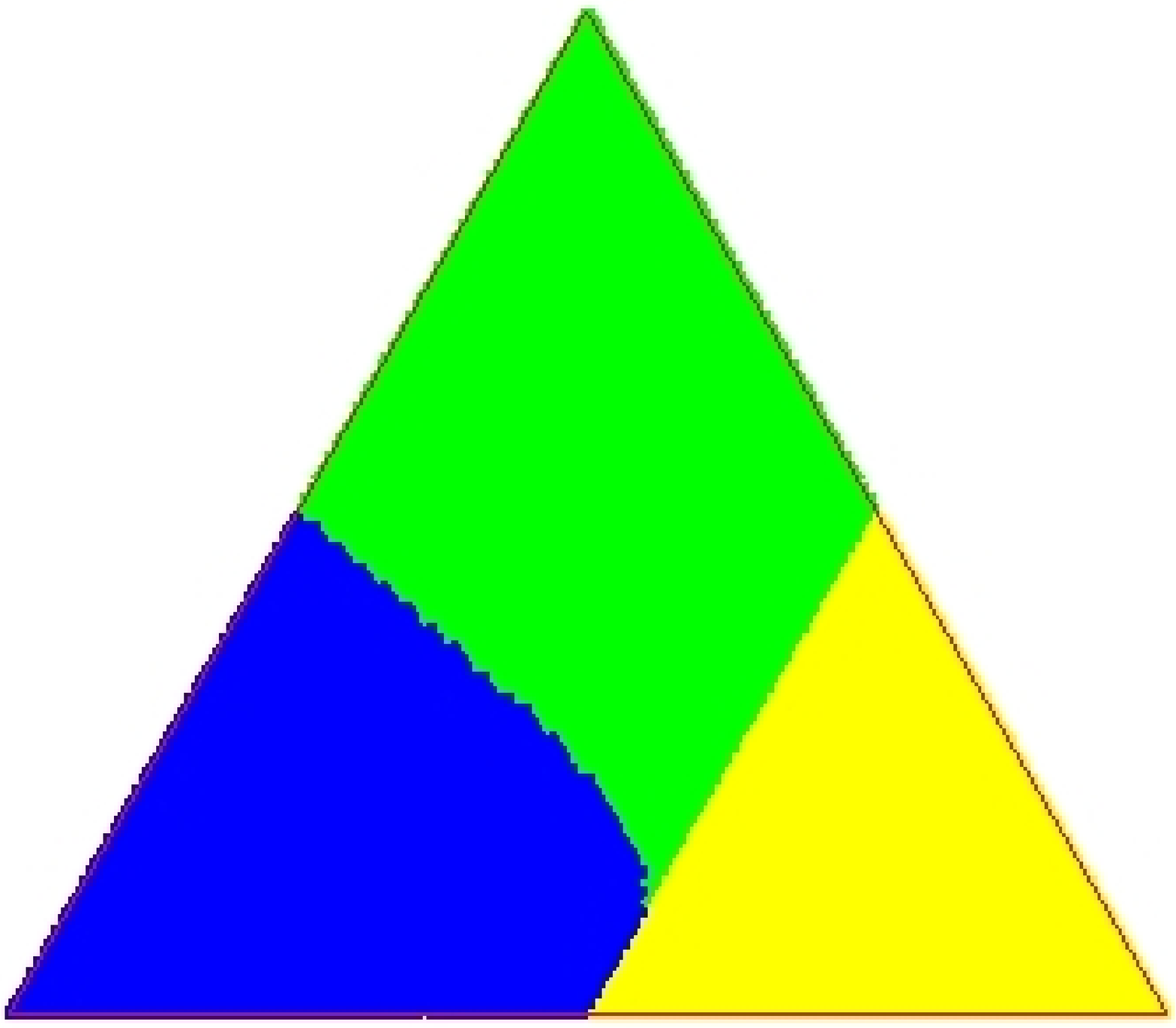}&
\includegraphics[width=2.6cm]{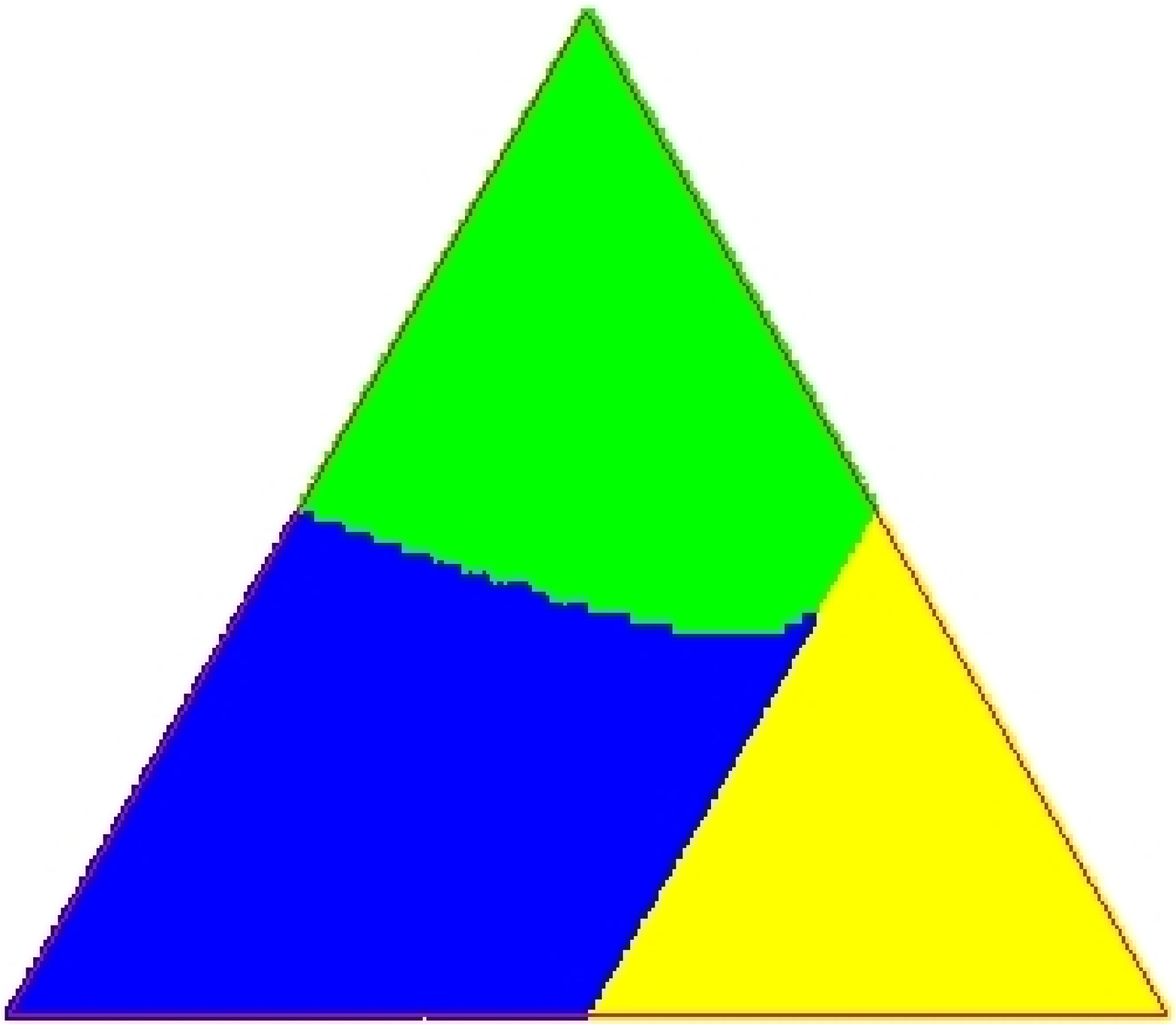}\\
\includegraphics[width=2.6cm]{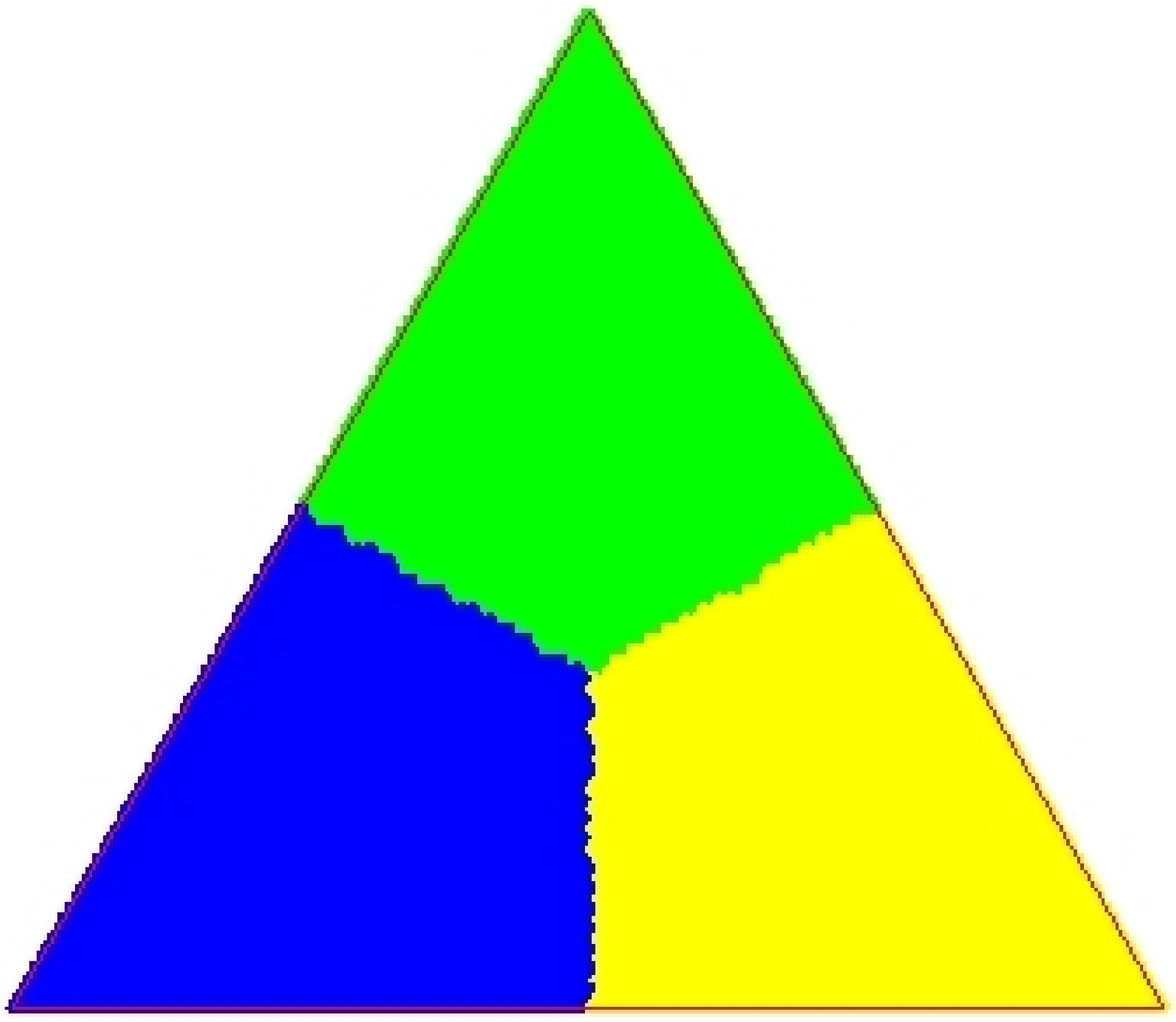}&
\includegraphics[width=2.6cm]{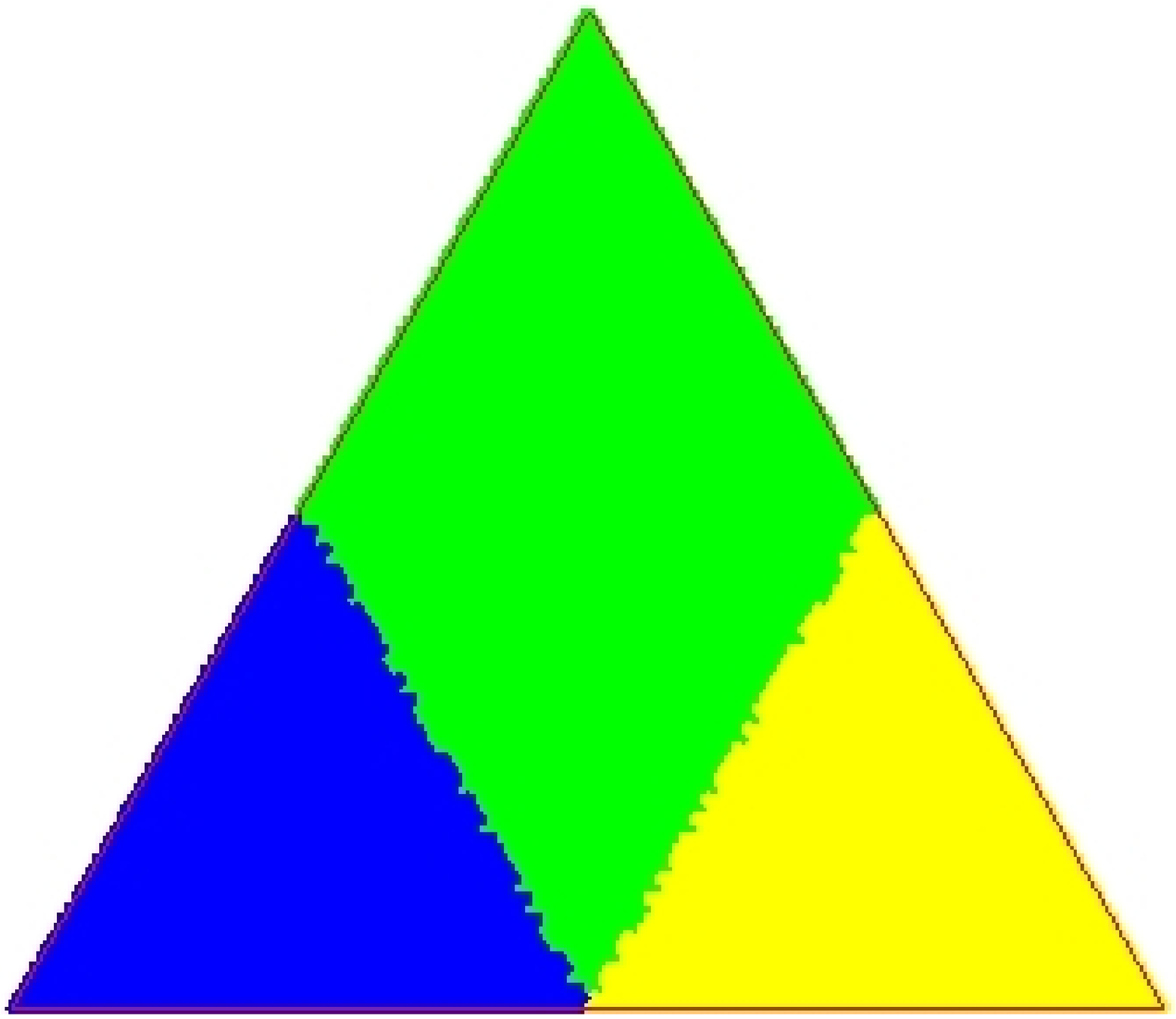}&
\includegraphics[width=2.6cm]{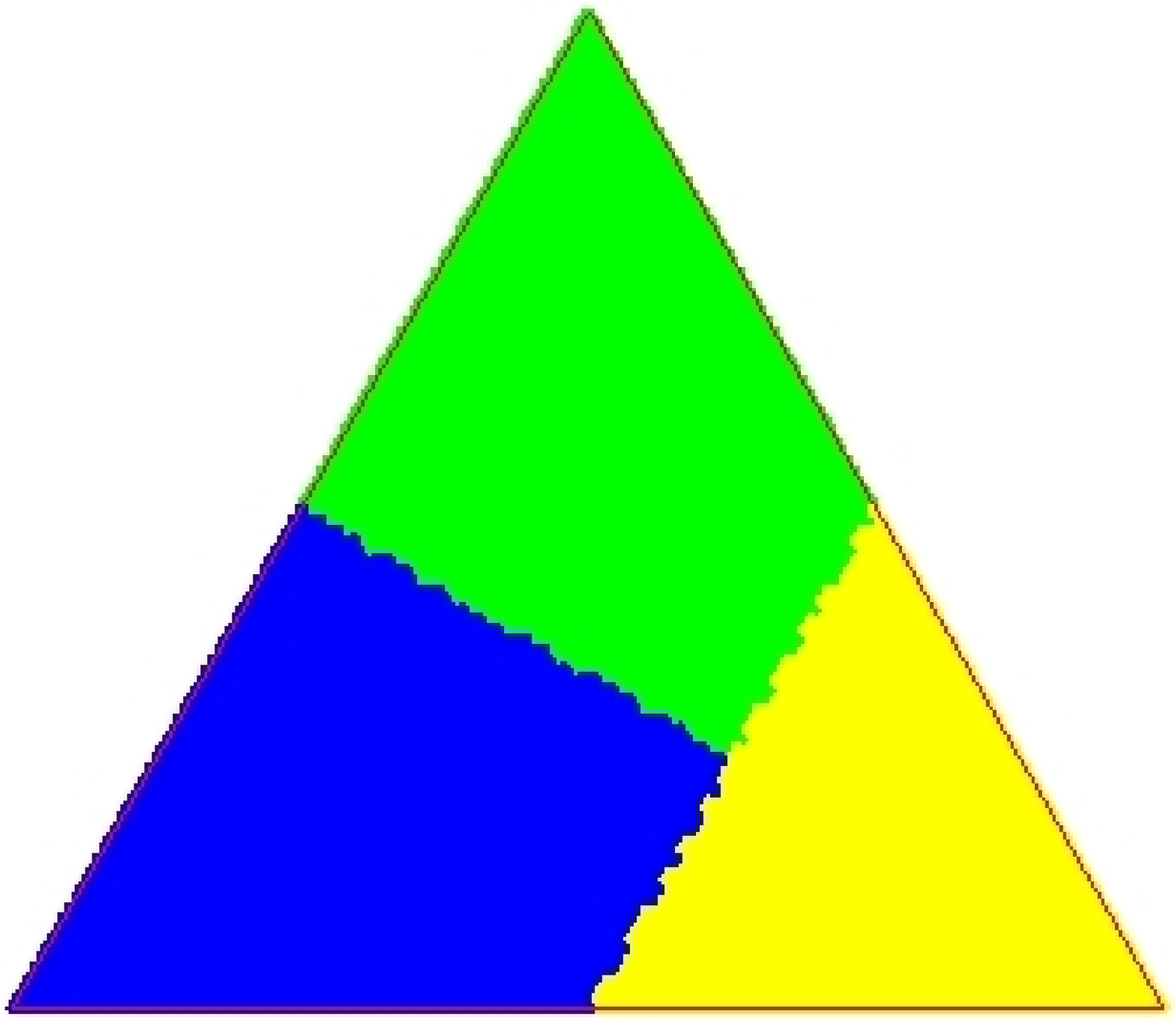}&
\includegraphics[width=2.6cm]{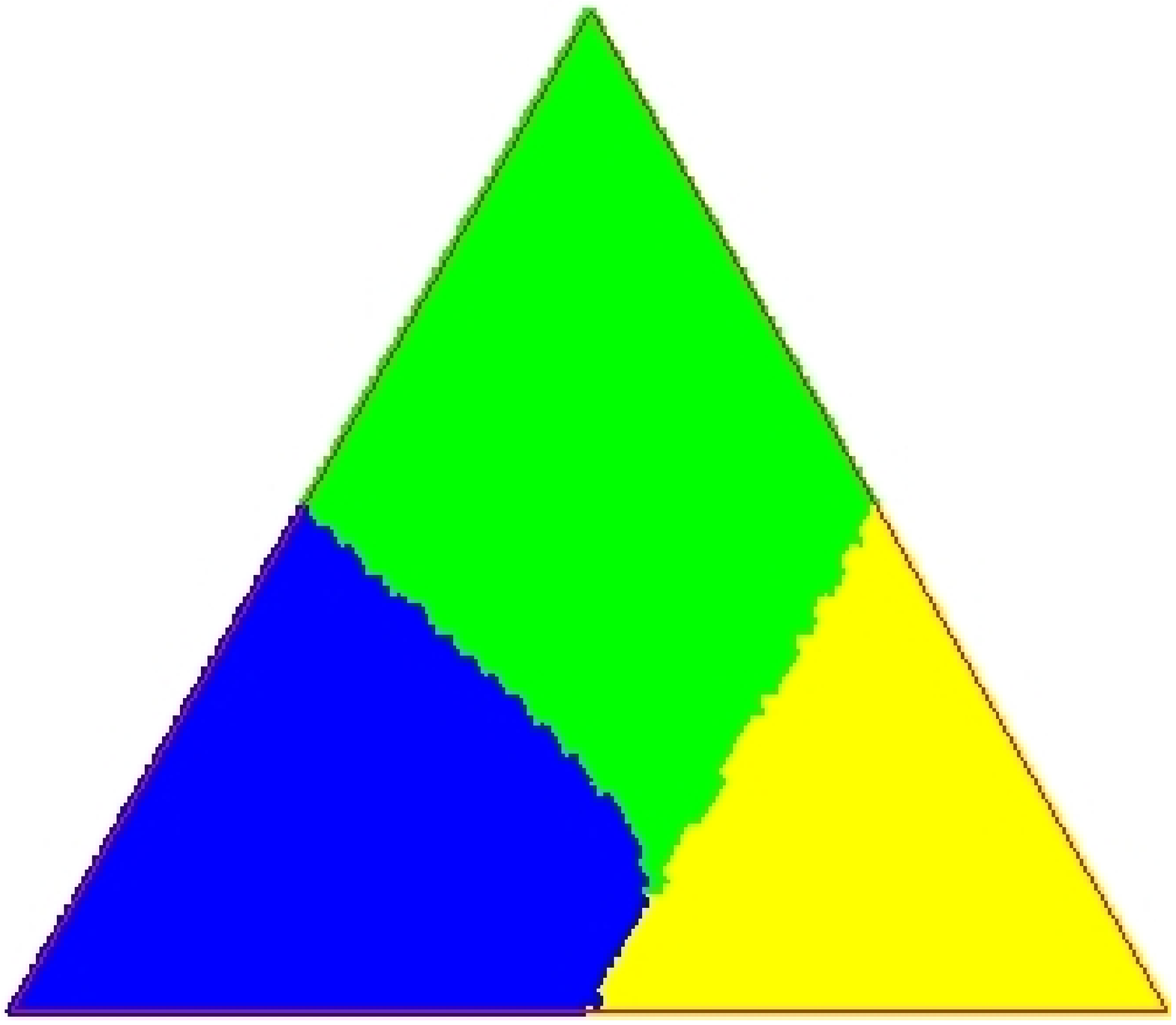}&
\includegraphics[width=2.6cm]{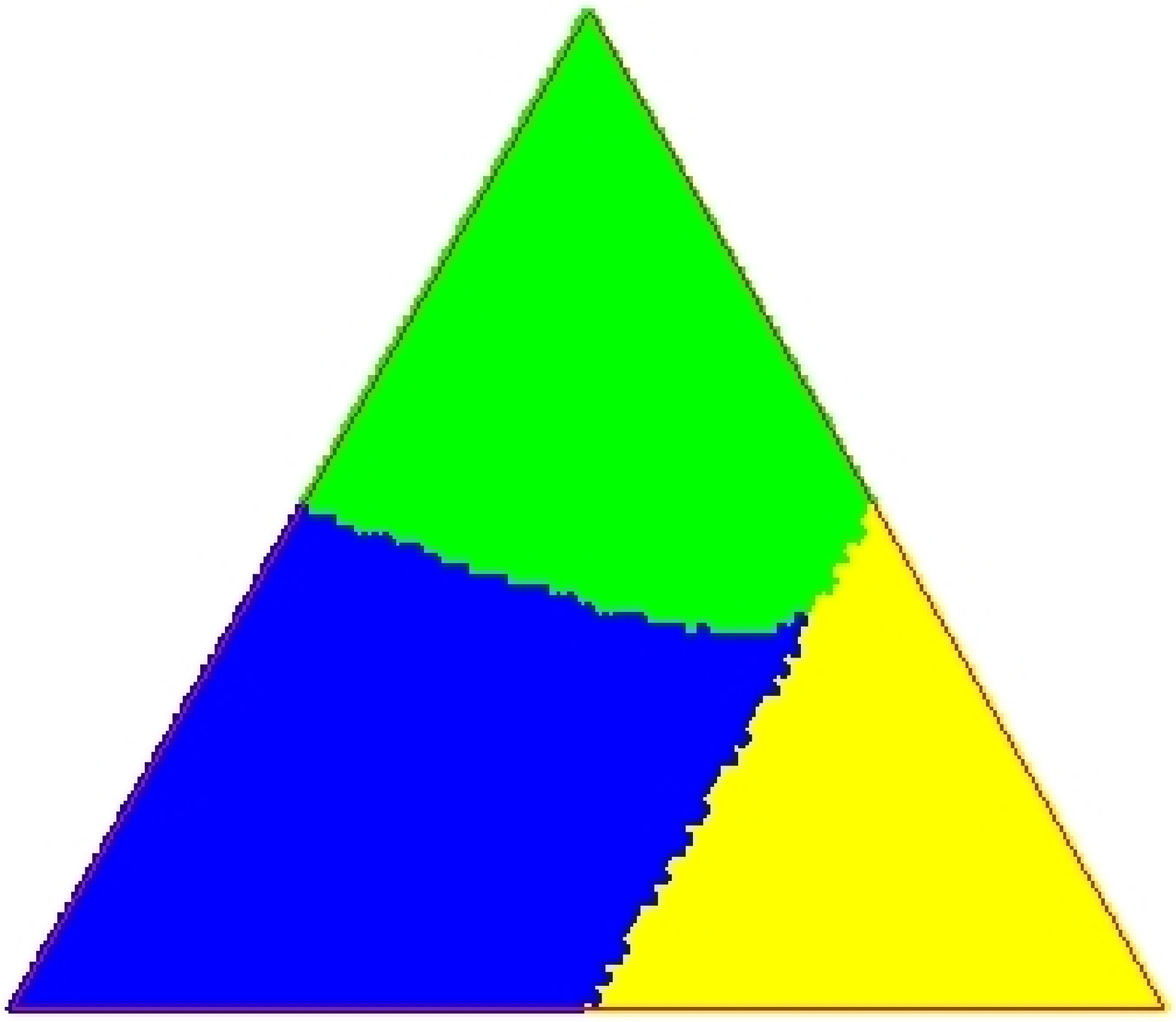}\\
\includegraphics[width=2.6cm]{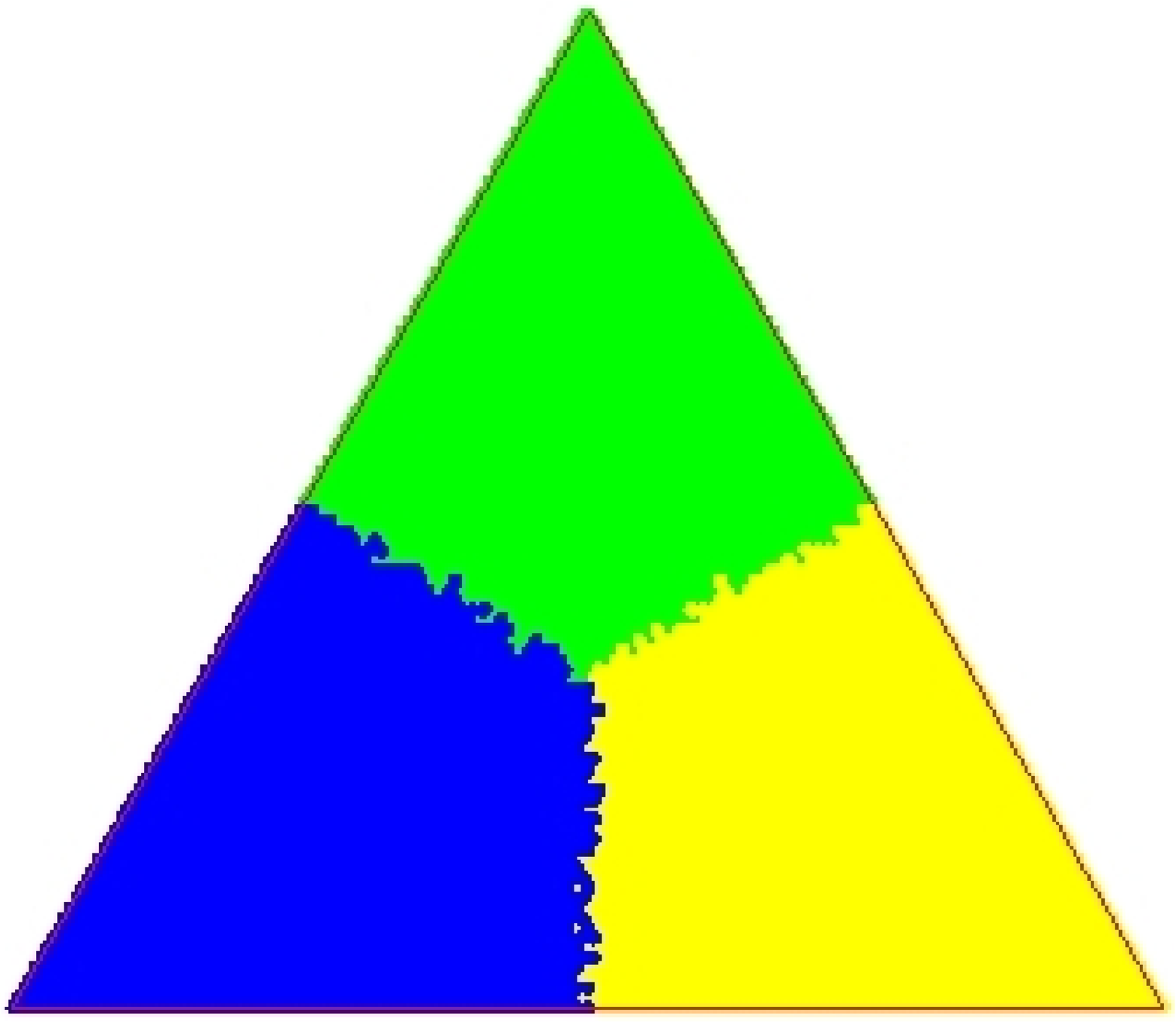}&
\includegraphics[width=2.6cm]{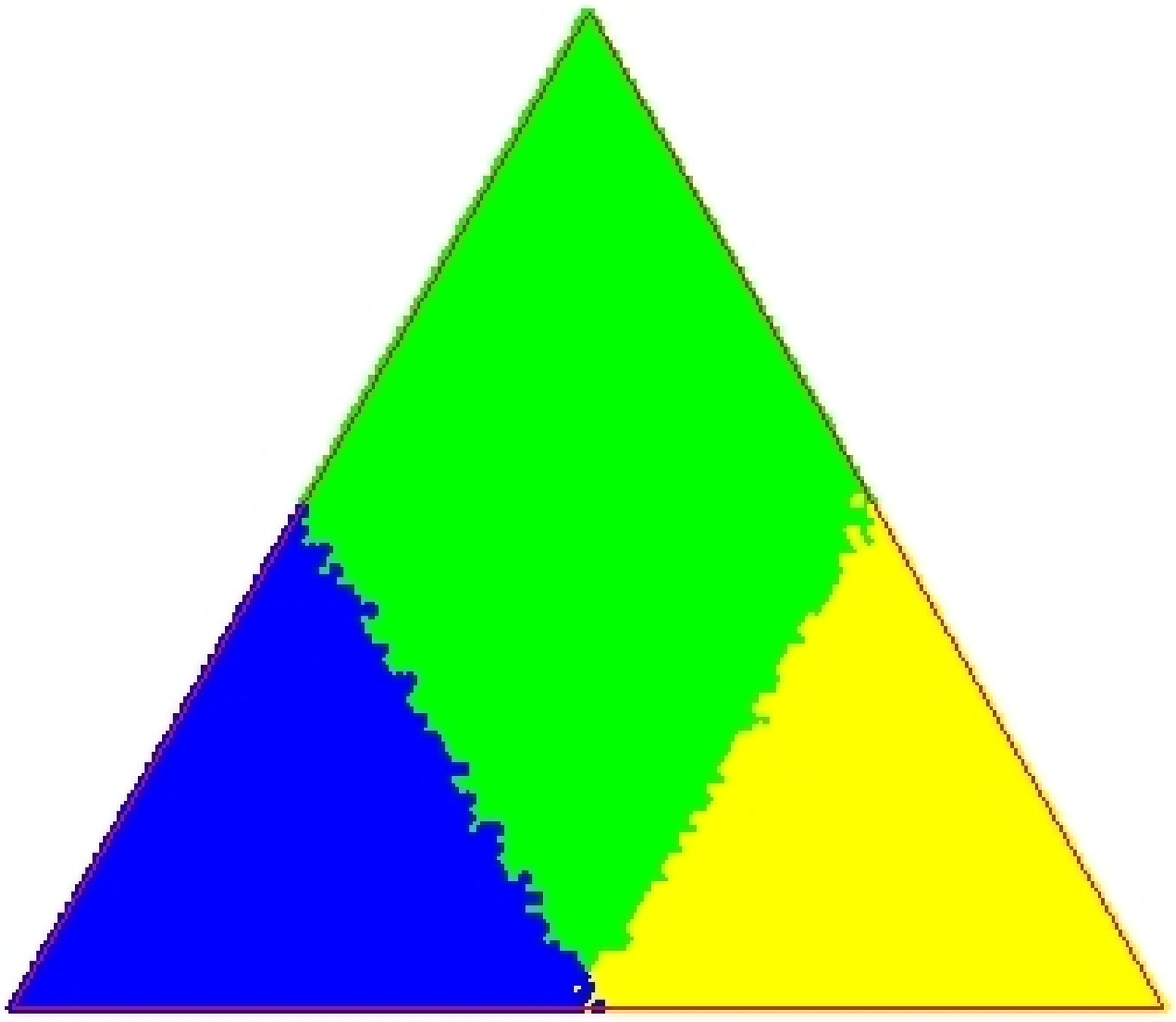}&
\includegraphics[width=2.6cm]{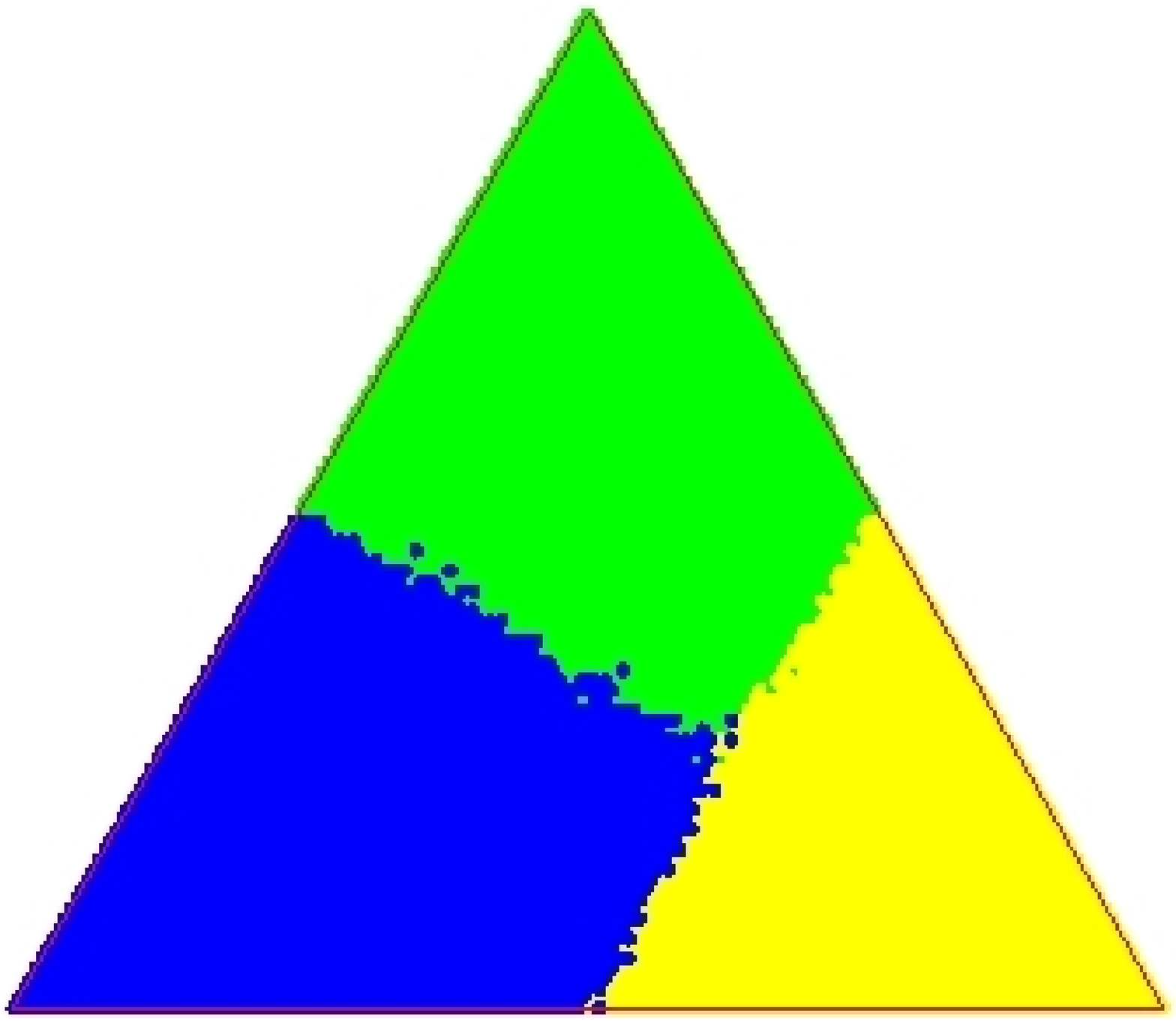}&
\includegraphics[width=2.6cm]{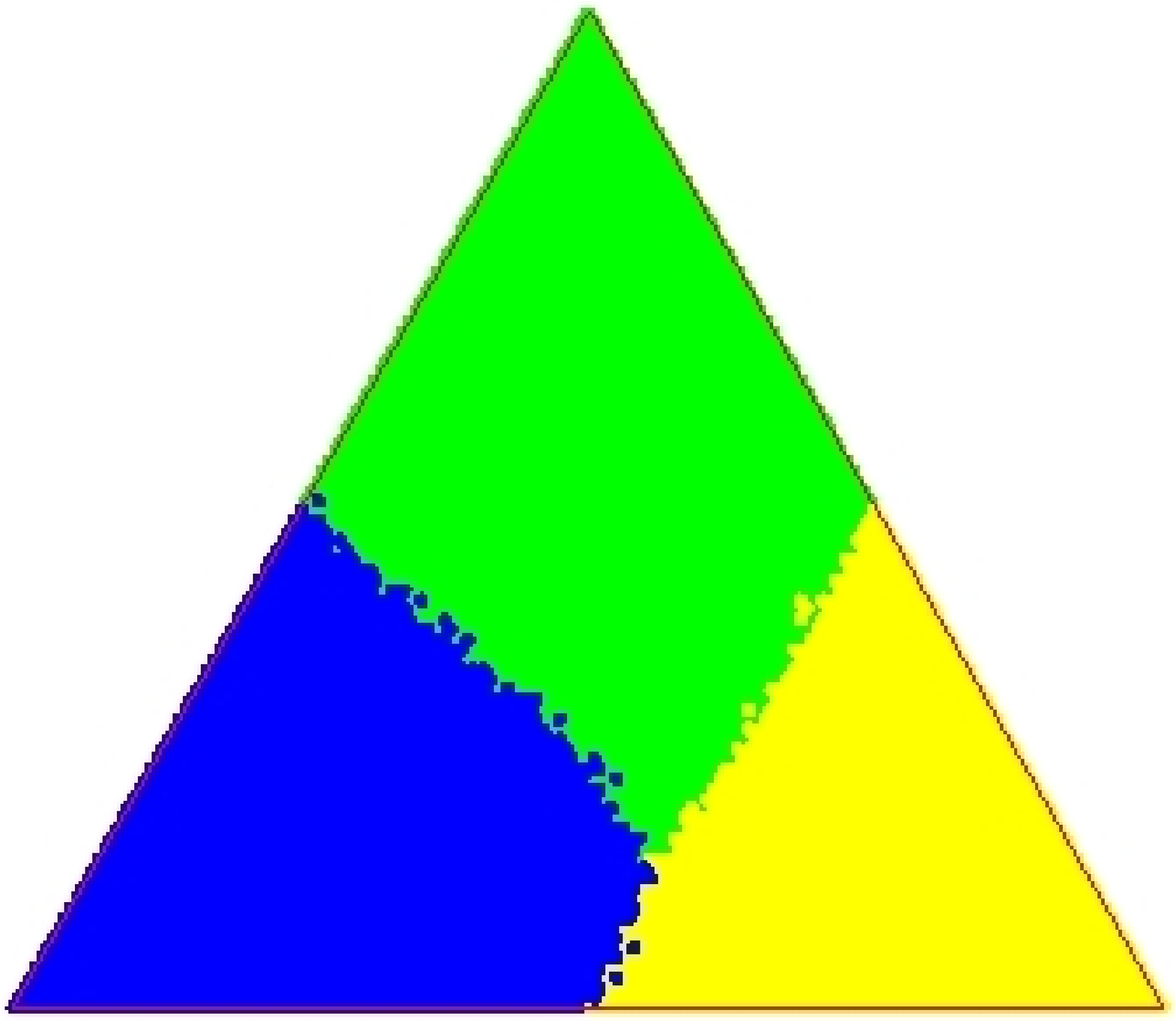}&
\includegraphics[width=2.6cm]{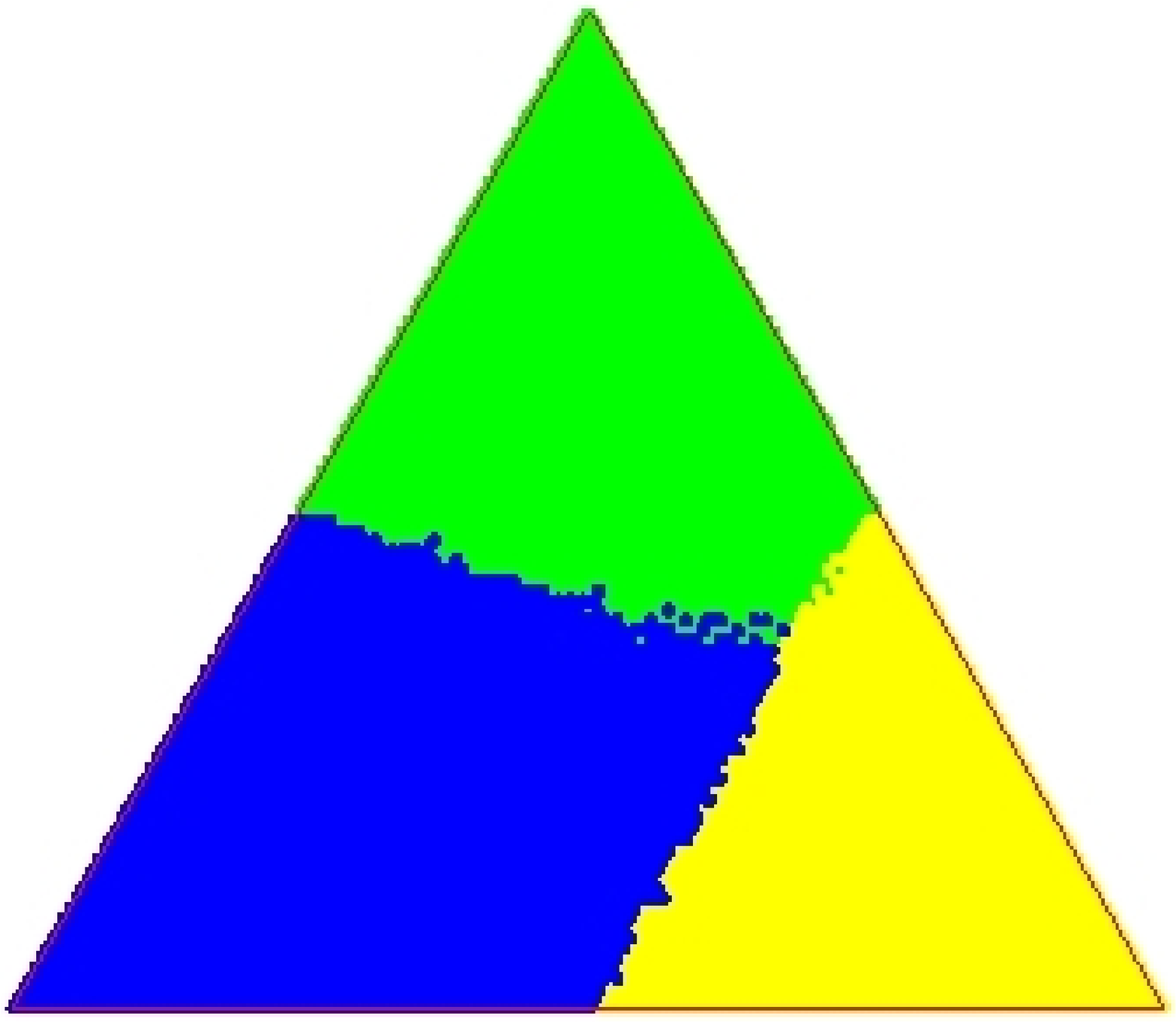}
\end{tabular}
\end{center}
\caption{Simulated phase diagrams for (from left to right)
$\alpha=\beta=\frac{1}{3}$; $\alpha=\beta=0$;
$\alpha=\frac{1}{2}$ and $\beta=0$; $\alpha=\frac{1}{3}$ and $\beta=0$;
$\alpha=\frac{2}{3}$ and
$\beta=0$.
From top to bottom system sizes of $N=10^4$, $N=2500$ and $N=100$.
The color of a point corresponds to the vertex
reached most often in 50 runs, starting from
that initial condition.
One can see how the boundary lines become blurred
for small systems.}\label{phd1:fig}
\end{figure*}

We now consider the behavior of systems with finite
values of $N$. Equations~(\ref{eva:eq},\ref{evb:eq}) provide
a good description of its behavior, provided
one decrees that complete polarization has
taken place whenever
\begin{equation}
p_i<\frac{1}{N},\qquad i\in\{\mathrm{A,B,C}\}.
\end{equation}
In this case, indeed, every single agent supports
the winning opinion. For $N$ large enough, the
attraction basins we have identified on the basis
of the iteration equations hold with practical
certainty, except very close to the boundaries.
When $N$ gets smaller, one can see some uncertainty
building up along the boundaries, as shown in
fig.~\ref{phd1:fig}. Here we assign a color to each
of the vertices of the triangle (black for A,
light grey for B, and dark grey for C) and we
assign to each of the initial conditions (a point
$(\pa^0,\pb^0)$ in the triangle) the color of
the vertex reached most often in 50 trials.
The diagram is evaluated on approximately 5000
different initial conditions with different
values of $N$: $N=10^4$, $N=2500$, and $N=100$.
It is clear that while the overall shape of the phase
diagram remains unchanged, the phase boundaries
become more and more blurred as $N$ decreases.

To get a more precise measure of the difference
between the results of simulations and the predictions
of the deterministic equations, we
compared the phase diagrams
one by one for all 5100 initial distributions and counted the
percentage of mismatches. Results are shown in
table~\ref{table:compSim}. One can see how the
quality of agreement diminishes with smaller system sizes,
but can be improved again by increasing
the number of runs over which the statistical average is taken.
This improvement only works up to a
certain level, though, the error margins for
5000 runs in a system with $N=100$ stay almost
exactly the same as for 1000 runs and are therefore not shown here.
As one can see in fig.~\ref{phd1:fig},
the mismatches occur only around the phase separation
lines. As soon as the initial distribution is
set a small distance apart from these lines, the
percentage of mismatch goes towards 0.
\begin{table*}
\begin{center}
\begin{tabular}{ccccccccc}
& \parbox{2cm}{\begin{center} $N=10^4$\\(50 runs)\end{center}}&
\parbox{2cm}{\begin{center} $N=2500$\\(50 runs)\end{center}}&
\parbox{2cm}{\begin{center}
$N=100$\\(50 runs)\end{center}}&
\parbox{2cm}{\begin{center} $N=100$\\(100 runs)\end{center}}&
\parbox{2cm}{\begin{center} $N=100$\\(1000
runs)\end{center}}\\
 \hline\\
\strut$\alpha=\frac{1}{3}$, $\beta=\frac{1}{3}$ & 0.7\% & 1.0\% & 1.8\%  & 1.5\% & 1.0\%\\
$\alpha=0$, $\beta=0$ & 0.1\% & 2.0\% & 2.4\% & 2.1\% & 2.1\%\\
$\alpha=\frac{1}{2}$, $\beta=0$ & 0.5\% & 1.5\% & 2.2\% & 1.9\% & 1.6\%\\
$\alpha=\frac{1}{3}$, $\beta=0$ & 0.1\% & 1.1\% & 2.0\% & 1.7\% & 1.3\%\\
$\alpha=\frac{2}{3}$, $\beta=0$ & 0.1\% & 1.2\% & 1.8\% & 1.6\% & 1.4\%
\end{tabular}
\end{center}
\caption{Percentage of mismatching points between
probability calculations and simulations for
different system sizes.}\label{table:compSim}
\end{table*}
As the error even for small systems with $N=100$ and only
50 runs is maximal 2.4\%, we conclude
that the deterministic eqs.~(\ref{eva:eq},\ref{evb:eq})
work very well for phase diagrams down to this size.

\subsection{Polarization time}
\begin{figure*}
\begin{center}
\includegraphics[width=4.5cm]{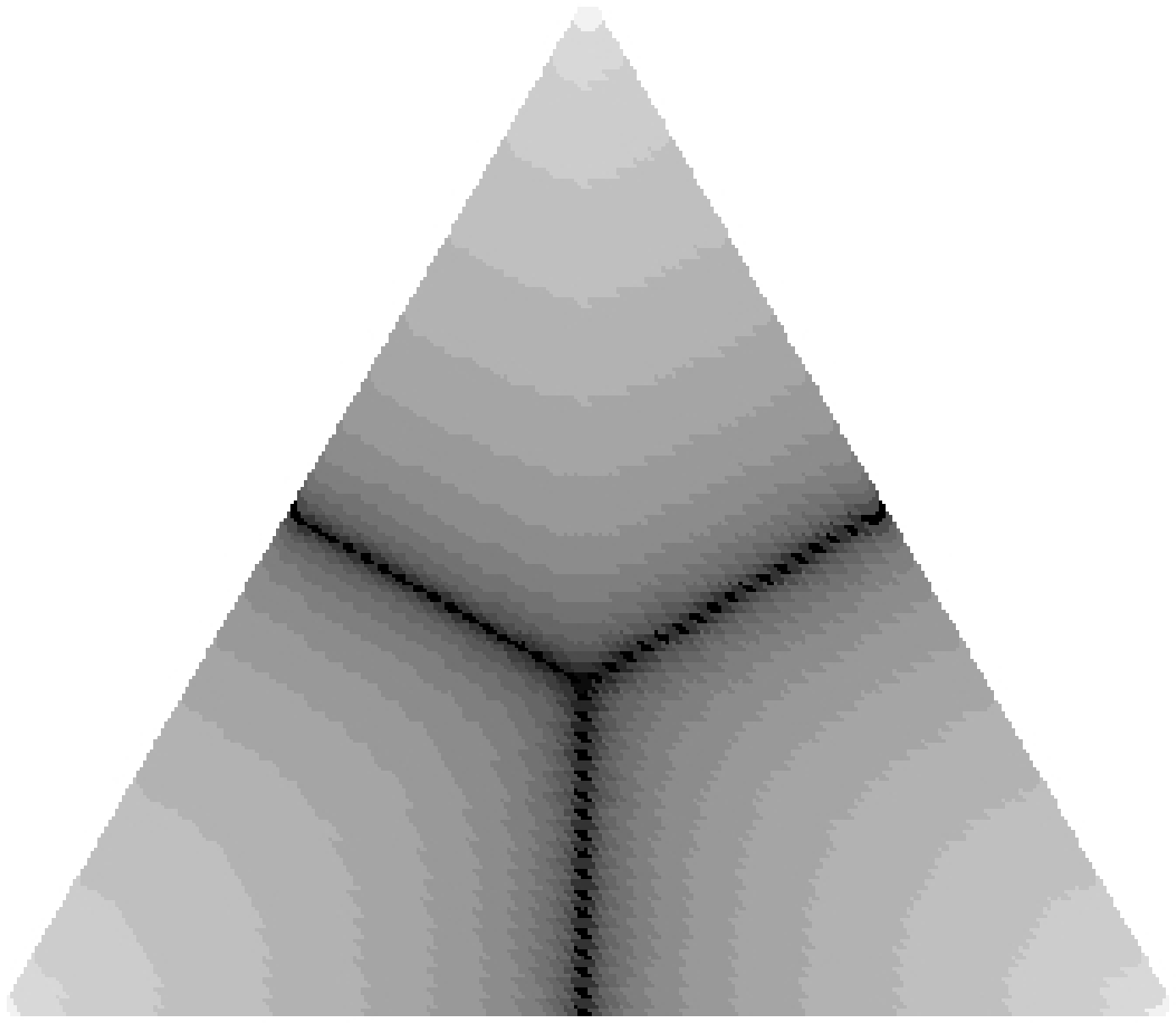}
\includegraphics[width=4.5cm]{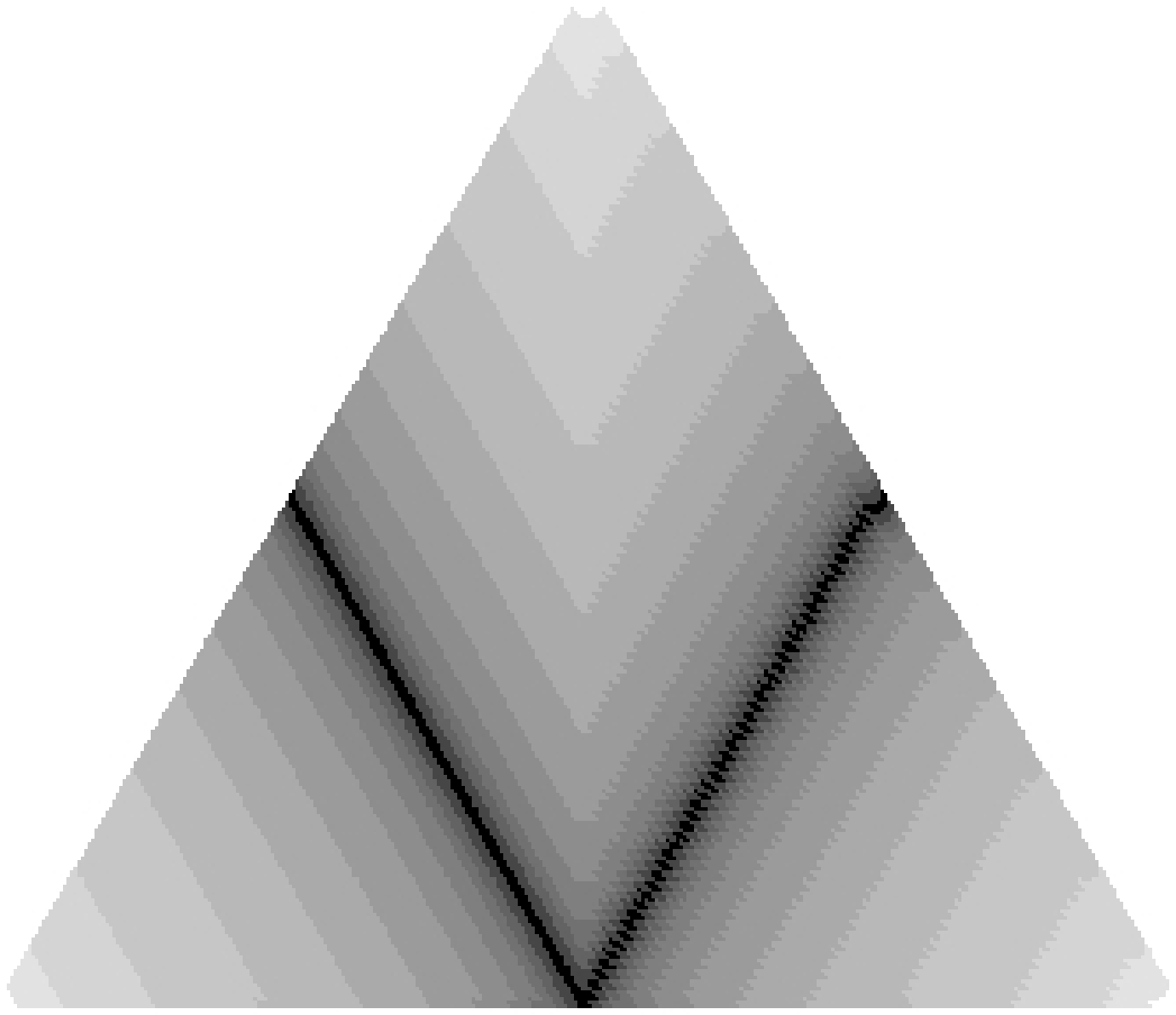}
\includegraphics[width=4.5cm]{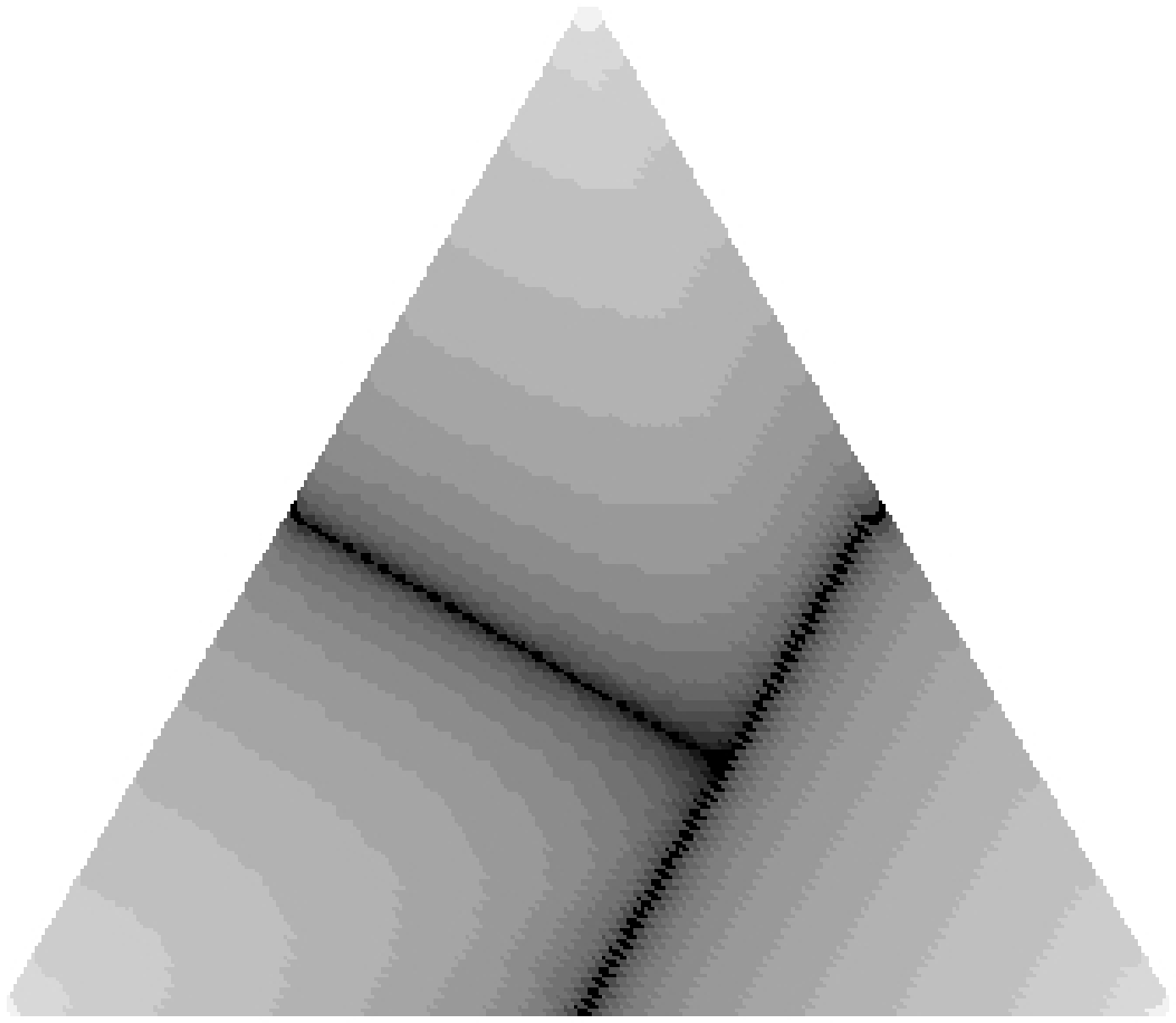}
\includegraphics[width=4.5cm]{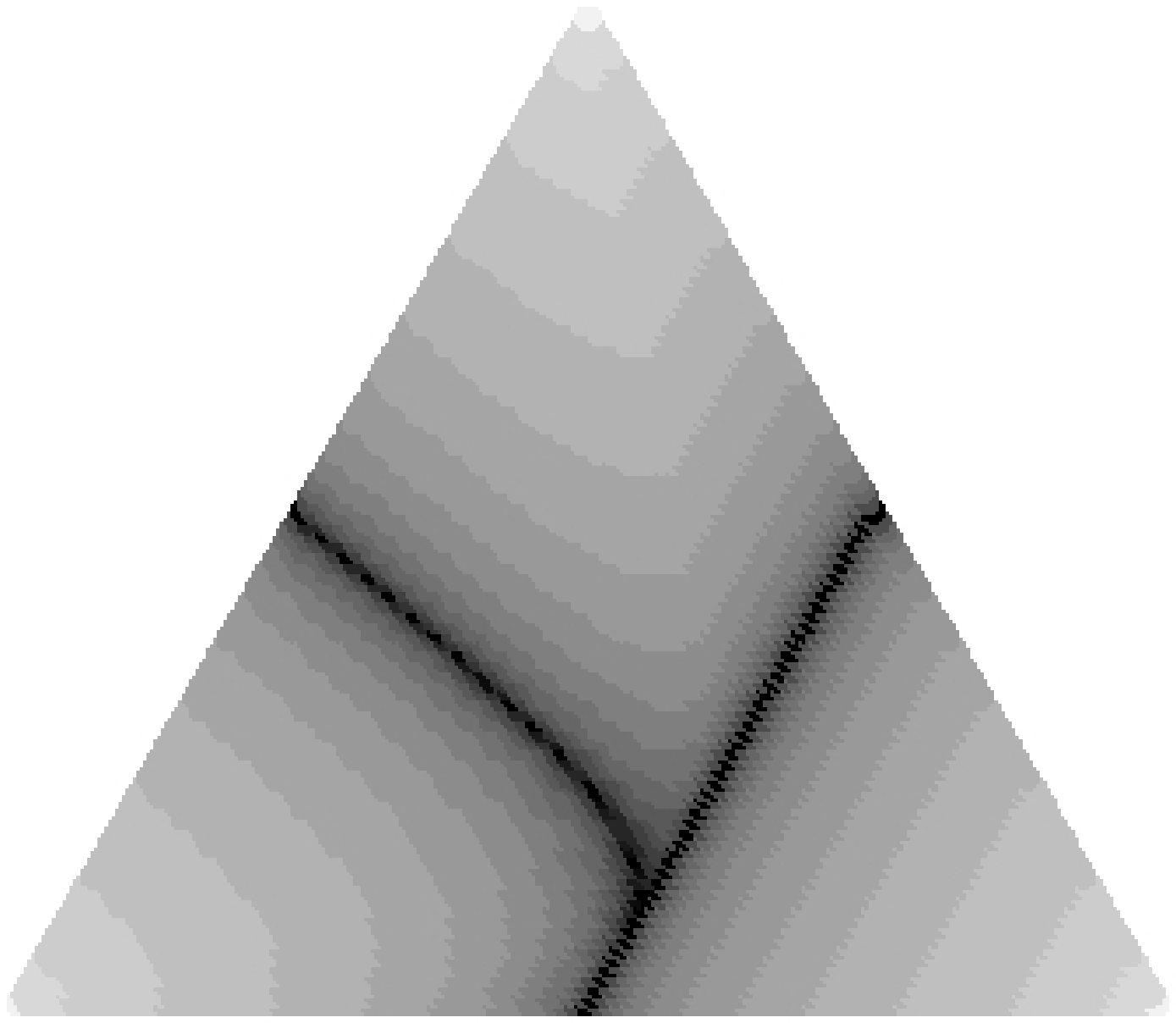}
\includegraphics[width=4.5cm]{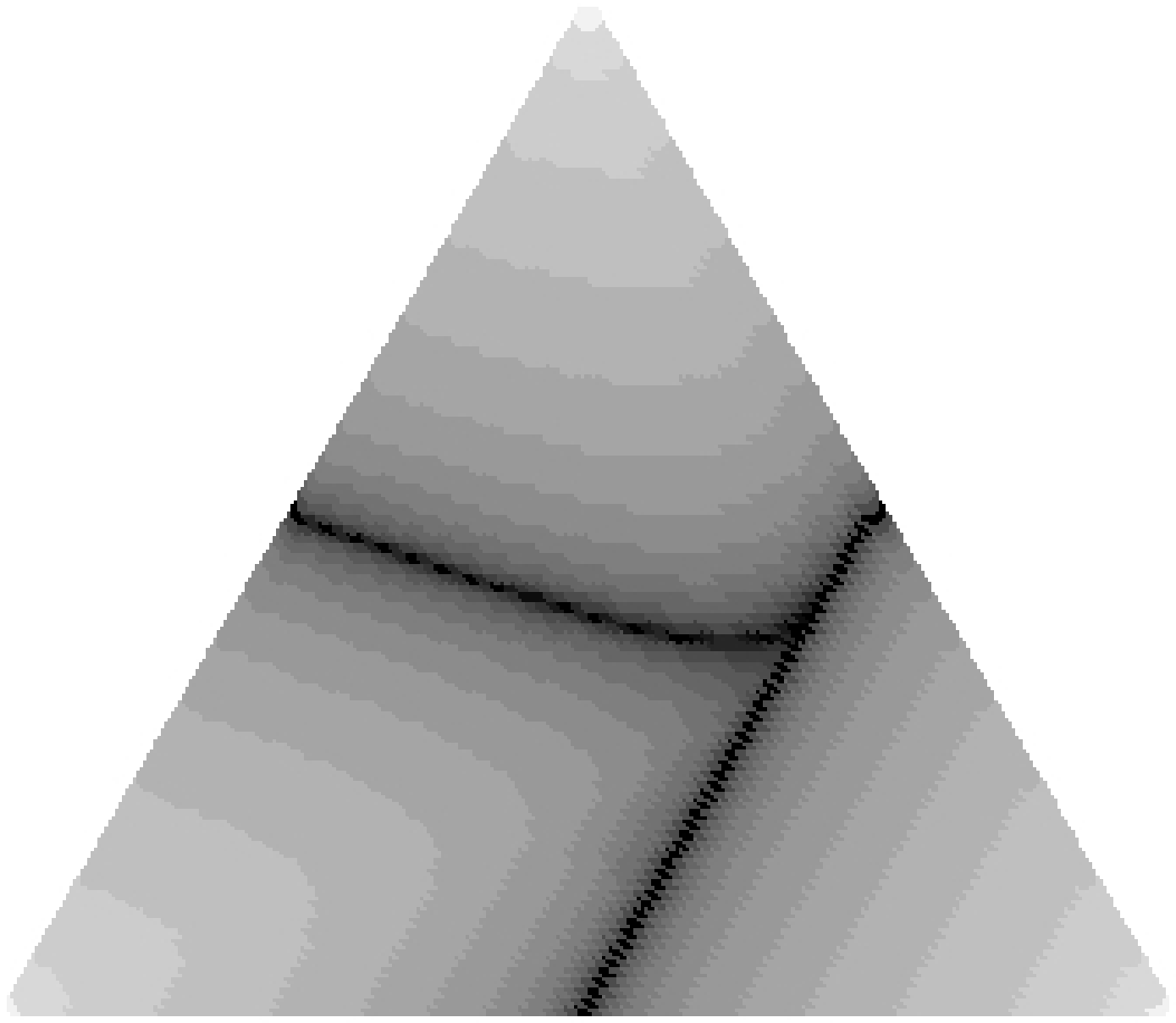}
\end{center}
\caption{Polarization time $T_\mathrm{p}$ in a system with
$N=10^4$ for different parameters obtained by
the deterministic evolution equations (\ref{eva:eq},\ref{evb:eq}).
Darker colors mean longer time. Note the peaks around the phase
boundaries. Above: left: $\alpha=\beta=\frac{1}{3}$; center:
$\alpha=\beta=0$; right: $\alpha=\frac{1}{2}$, $\beta=0$. Below: left: $\alpha=\frac{1}{3}$,
$\beta=0$; right: $\alpha=\frac{2}{3}$, $\beta=0$.}\label{fig:timeTriangles}
\end{figure*}

\begin{figure}
\begin{center}
\psfrag{0}[r]{$\alpha=\beta=0$}
\psfrag{3}[r]{$\alpha=\beta=1/3$}
\includegraphics[width=9cm]{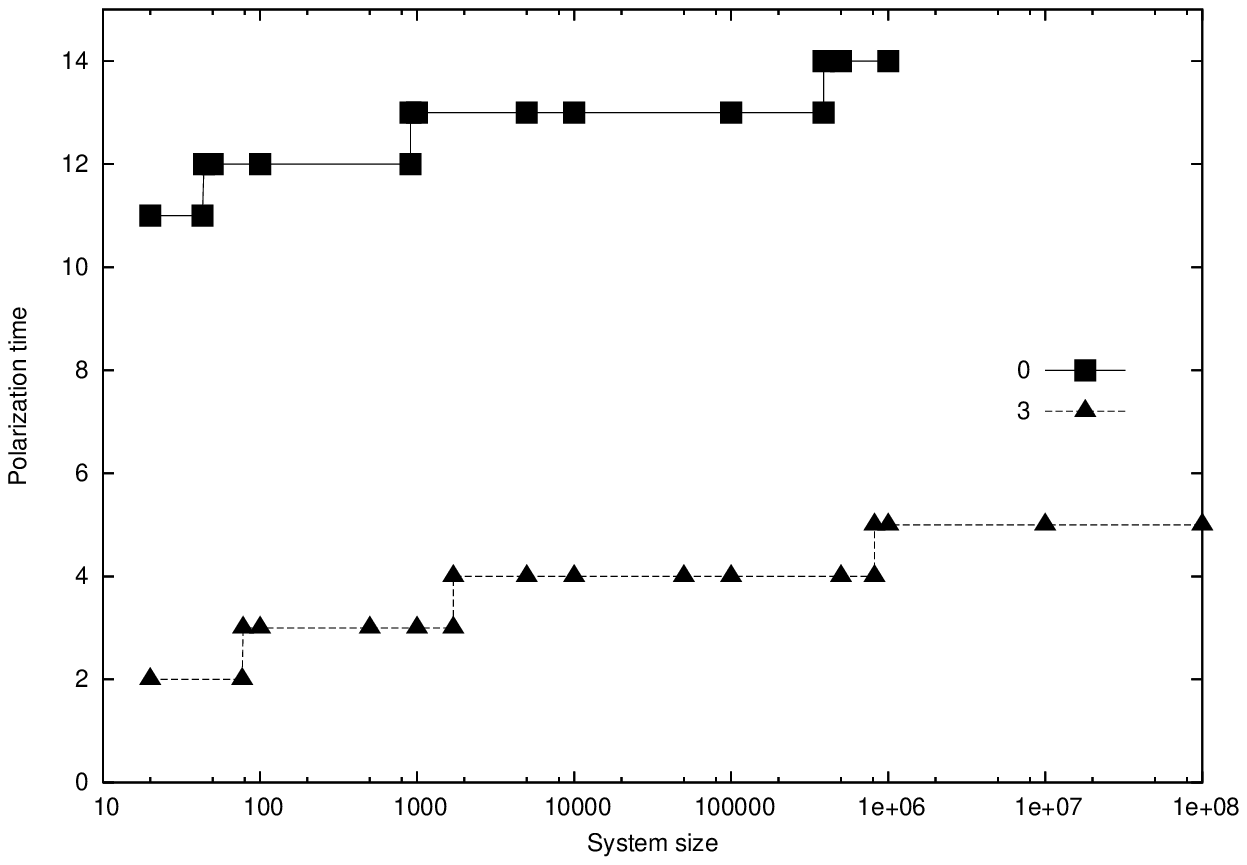}
\end{center}
\caption{Polarization times $T_\mathrm{p}$ obtained by iteration of eqs.~(\ref{eva:eq},\ref{evb:eq}) for different
system sizes.
The system starts fom $\pa=0.51$, $\pb=0.49$ for the case $\alpha=\beta=0$ (squares) and from $\pa=0.1$, $\pb=0.1$ for $\alpha=\beta=1/3$ (triangles). The same convention applies to figs.~\ref{fig:timeSimulated} and \ref{fig:timeRelError}.
The system starting far from the triple point G (lower curve) converges a lot
faster on a polarization. Even over a wide range of system sizes $T_\mathrm{p}$ remains virtually constant.
}\label{fig:timeAnalytical}
\end{figure}

It is also interesting to discuss the polarization time $T_\mathrm{p}$,
i.e., the number of discussion cycles needed
to reach complete polarization. It depends on the initial condition
as well as $\alpha$ and $\beta$. We list in table \ref{table:times}
polarization times for
a few interesting configurations.
\begin{table*}
\begin{center}
\begin{tabular}{ccccccc}
$\alpha$&$\beta$&$p_A$&$p_B$&winner&\parbox{2cm}{\begin{center}time
$T_\mathrm{p}$\\(probabilities)\end{center}}&
\parbox{2cm}{\begin{center}time $T_\mathrm{p}$\\(simulated)\end{center}}\\
\hline\\
0 & 0 & 1    & 0    & A & 0  & 0 $\pm$ 0\\
0 & 0 & 0.51 & 0.49 & A & 13 & 13.1 $\pm$ 1.4\\
0 & 0 & 0.49 & 0.49 & C & 13 & 13.8 $\pm$ 1.1\\
0 & 0 & 0.30 & 0.30 & C & 5  & 5.4 $\pm$ 0.5\\
0 & 0 & 0.1  & 0.1  & C & 3  & 3.1 $\pm$ 0.3\\
1/3 & 1/3 & 0.1  & 0.1  & C & 4 & 4.0 $\pm$ 0.3\\
1/2 & 0   & 0.1  & 0.1  & C & 4  & 4.0 $\pm$ 0.2\\
1/2 & 0   & 0.3  & 0.51 & B & 13 & 13.2 $\pm$ 1.4
\end{tabular}
\end{center}
\caption{Polarization time $T_\mathrm{p}$ for a few interesting configurations.
Results obtained by
the evolution equations (\ref{eva:eq},\ref{evb:eq})
and by simulations for a system of $N=10^4$.
Simulated results averaged over 100
runs.}\label{table:times}
\end{table*}
In fig.~\ref{fig:timeTriangles} the polarization
times are displayed as a function
of the initial condition over the whole
triangle. One can see that the polarization time becomes longer
as the initial distributions approaches the phase
boundaries.
The scaling of $T_\mathrm{p}$ with system size $N$,
using the deterministic evolution equations,  is shown in fig.~\ref{fig:timeAnalytical}
for two configurations, one with an initial distribution close to the
triple point G with $\alpha=\beta=0$ starting from $\pa=0.51$ and $\pb=0.49$, the other
far from the triple point with $\alpha=\beta=\frac{1}{3}$
starting from $\pa=\pb=0.1$.
One sees that, even if $N$ varies  from 25 to $10^8$,
$T_\mathrm{p}$ remains virtually constant,
varying only between 11 and 14 in the first and between 2 and 5
in the second case.

\begin{figure}
\begin{center}
\psfrag{0}[r]{$\alpha=\beta=0$}
\psfrag{3}[r]{$\alpha=\beta=1/3$}
\includegraphics[width=9cm]{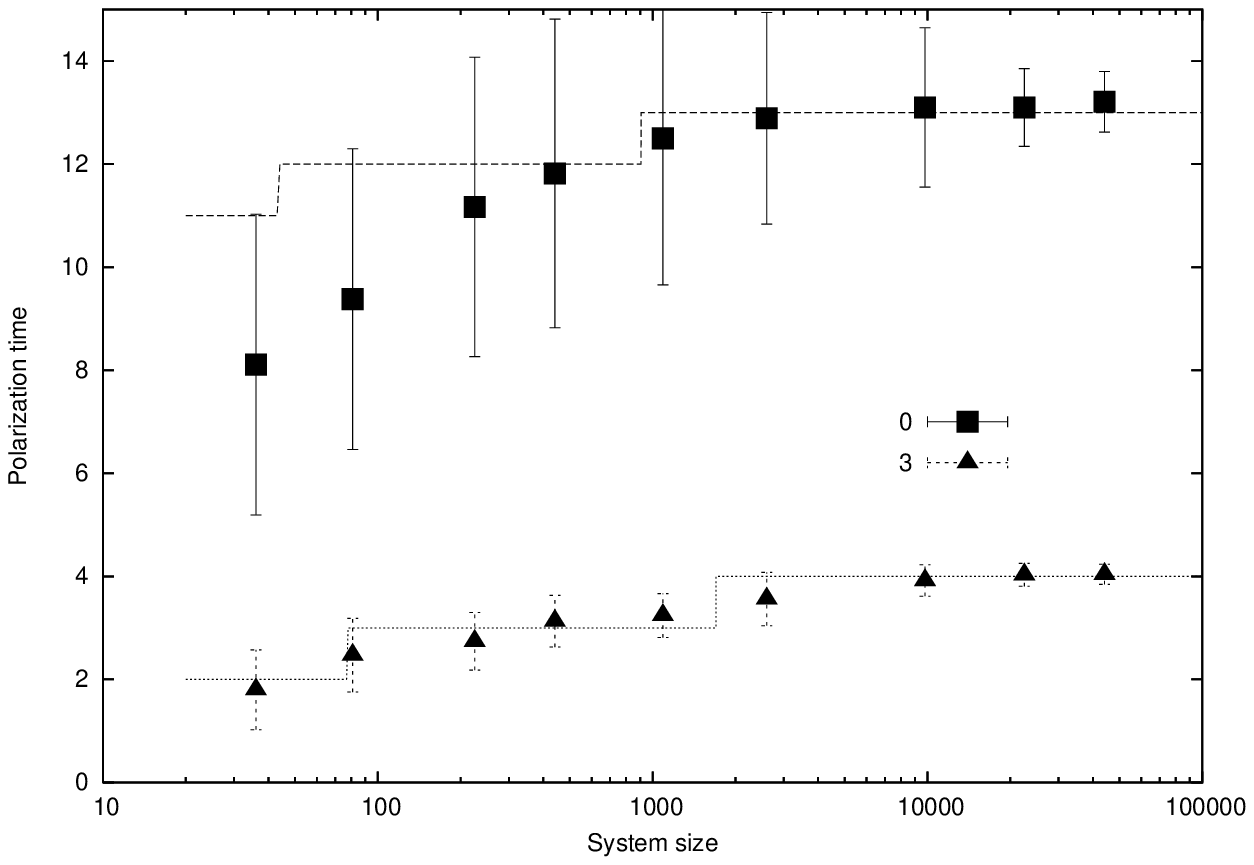}
\end{center}
\caption{Polarization times $T_\mathrm{p}$ obtained
through simulations for different system sizes. As
expected, the error bars become smaller with increasing system size.
Overall, error bars for the
system starting far from the triple point
G are much smaller than for the other system.
Results averaged over 100 runs. }\label{fig:timeSimulated}
\end{figure}

The scaling of the simulated polarization time for
the same configurations as above is
shown in fig.~\ref{fig:timeSimulated}.
The picture now looks quite different than for the phase diagrams.
Regarding $T_\mathrm{p}$ only
for systems with about 1000 agents the agreement is still acceptable,
although the error bars are
already quite large.
Overall the agreement for the system starting
far from the triple point is much better
than for the system starting close to it.

\begin{figure}
\begin{center}
\psfrag{0}[r]{$\alpha=\beta=0$}
\psfrag{3}[r]{$\alpha=\beta=1/3$}
\includegraphics[width=9cm]{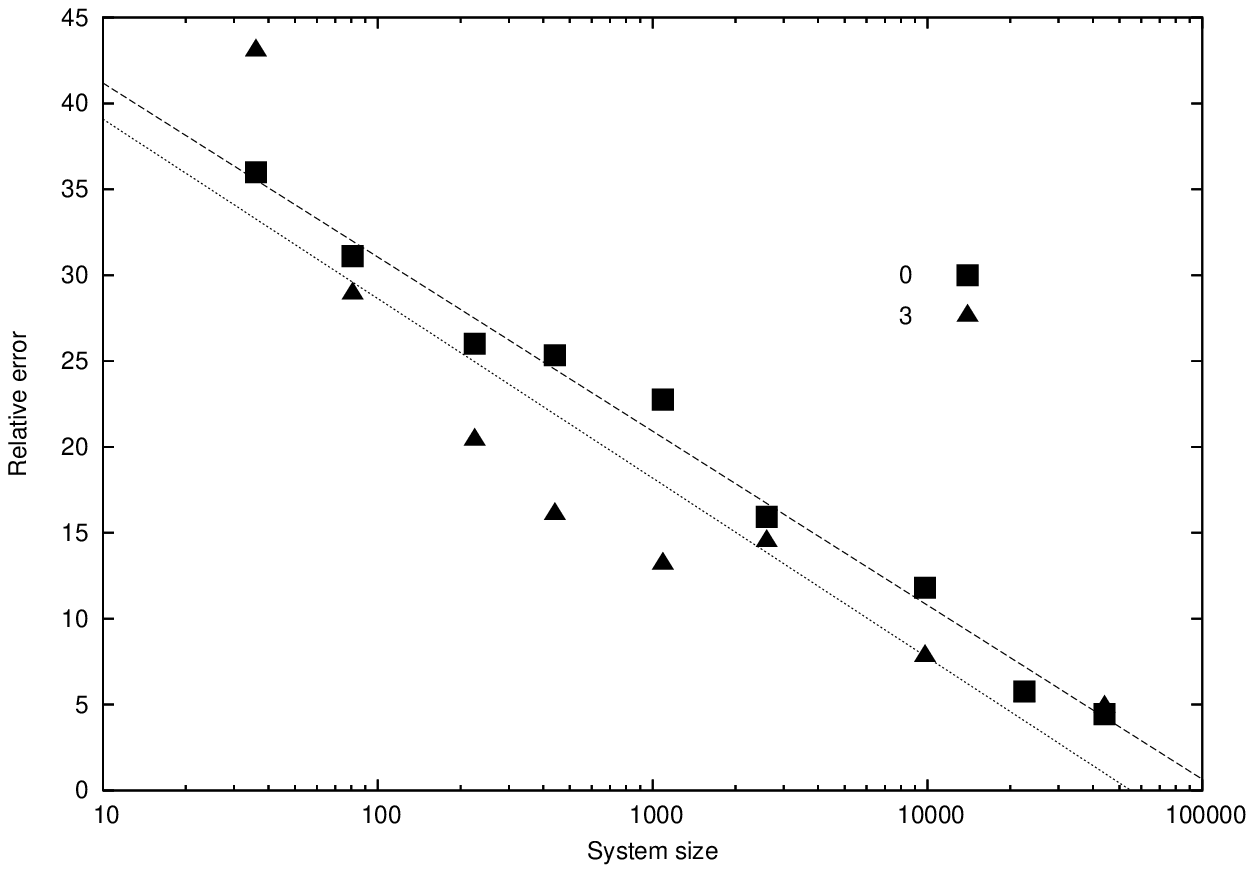}
\end{center}
\caption{Relative error of the simulated polarization time $T_\mathrm{p}$.
For systems with $N<1000$ the
error becomes very large and predictions seem unreliable.}\label{fig:timeRelError}
\end{figure}

Looking at the relative error shown in fig.~\ref{fig:timeRelError},
one must suggest that for small
systems with $N<1000$ a prediction of the
polarization time via the evolution equations
(\ref{eva:eq},\ref{evb:eq}) is unreliable.
However, predictions about the winning opinion remain
fairly reliable as described in the previous section.

\section{Concluding remarks}
\label{concl:sec}
In this paper we presented an extension of Galam's model
for a two-choice system \cite{Gal02} to a
system in which each agent can choose between three possible opinions.
All research has been done
for discussion groups of size 3 and in a randomly-localized model.
To resolve ties, two
new parameters $\alpha$ and $\beta$ measuring ``argument strength'' were
introduced. Opinion distributions could be conveniently visualized in an equilateral triangle.

We showed that, independent of the parameter values, the system will almost always reach a complete
polarization after a small number of discussion cycles.
These polarization times are virtually independent of system size.
We have drawn the phase diagrams, associating to each initial condition $(\pa^0,\pb^0)$
the opinion (A, B, or C) which eventually wins.
Finally, a number of simulations has been conducted to show that the deterministic evolution equations used to
obtain the above results are reliable down to systems with approximately 100 agents, as far as the
phase diagram is concerned, and to $N\simeq 1000$ for the polarization times.

Similar phase diagrams are obtained with a serial update dynamics, in
which, at each microscopic time step, three different agents
are picked up at random and their opinions updated according
to the local majority rule described above. One counts
a time step when each individual has taken part to
three discussions on average. The polarization times
with this dynamics are longer. Indeed, while in the parallel
dynamics discussed so far the approach to the stable fixed points is ultrafast
(the deviation gets squared at each time step) it is only
exponential in the serial update dynamics. Thus
\begin{equation}
T_\mathrm{p}\sim\cases{\sqrt{\ln N},& with parallel dynamics;\cr
\ln N,& with serial dynamics.}
\end{equation}

Our results show that dealing with three opinions is much more
complex than with two. From the complexity of the various flows,
it is seen that initial conditions are not enough to predict
any outcomes since the values of the parameters $\alpha$ and
$\beta$ are instrumental to determine the highly nonlinear
road to success or failure in the rather quick competition in the
opinion forming process. What appears as a successful strategy for
one opinion in the first steps may turn out to be a total disaster
at the end. At this stage of our investigation the following clues can be outlined.
\begin{enumerate}
\item It seems that we have here some hint on why the various
attempts to create a third ``way''  in democratic countries
have always failed. Starting from an A-B two-party situation two
possibilities exist for the creation of a third one C.
Either it is created on its own against A and B or it is
``helped'' by one of the two  big forces, say A. In the first
case we have $\alpha=\beta=\frac{1}{3}$ making C much further
from the boundaries since by its nature it is a new comer.
Therefore it will rather quickly disappear. In the second
case we have $\alpha=\beta=0$ since at a tie, A goes for C.
In this case C will start growing slowly at the expense of both B and A.
To stop it from spreading all over the unique solution is to have A and B
  to make a coalition against C with $\alpha=\beta=\frac{1}{2}$.
\item Focusing on the polarization effect with one opinion spreading
in two opinion competitions, it seems that the best strategy is not
 to reinforce its own side but instead to create a third opinion
 which cooperates with either one of the others. It makes the
 new very minority opinion to eventually  win the competition.
 \end{enumerate}
The above ideas are of course qualitative and need to be deepened.
This paper is just a beginning outlining some basic features of
 three-opinion systems. There is much room for further research,
 as for example the introduction of a geographical distribution and
 the influence of neighborhood relations \cite{Tessone}. Another
 interesting topic would be the behavior of discussion groups
 with more than three members \cite{Gal91}. To introduce some
 dependence of the parameters $\alpha$ and $\beta$ on the values of
 the respective support for A and B would be also of a great interest.

\section*{Acknowledgments}
We would like to thank the COST organization
for a scholarship to support S. Gekle's stay
in Paris, where this paper was written. S. Galam
is grateful to INFM for supporting his trip
to Naples, where this collaboration was started.

\end{document}